\documentclass[final,5p,twocolumn,numbers]{elsarticle} 

\usepackage[english]{babel}
\usepackage{amsmath}
\usepackage{amssymb}
\usepackage{newtxtext}   
\usepackage{newtxmath}   
\usepackage{graphicx}
\usepackage{dcolumn}
\usepackage{bm}
\usepackage{enumitem}
\usepackage{booktabs}
\usepackage{hyperref}
\hypersetup{
	colorlinks=true,
	linkcolor=blue,
	urlcolor=blue,
	citecolor=blue,
	pdftitle={Bifurcation and Multistability in a Collective-Risk Public Goods Game with Quorum-Activated Protection},
	pdfauthor={Shulan Li, Xiaogang Li, Jun he, Lei Shi, Attila Szolnoki} 
}

\begin{document}
	\raggedbottom
	\setlength{\parskip}{0pt}
	\begin{frontmatter}

		\title{Bifurcation and Multistability in a Collective-Risk Public Goods Game with Quorum-Activated Protection}
		
		\author[inst1]{Shulan Li}
		\author[inst2]{Xiaogang Li}
		\author[inst2]{Lei Shi\corref{cor1}}
		\author[inst3]{Jun He\corref{cor1}}
		\author[inst4]{Attila Szolnoki}
		
		\cortext[cor1]{~Corresponding authors.\\\hspace*{1.8em}\raggedright 
			\textit{Email addresses:} \href{mailto:shi\_lei65@hotmail.com}{shi\_lei65@hotmail.com} (Lei Shi), \href{mailto:hejun@kmu.edu.cn}{hejun@kmu.edu.cn} (Jun He)}
		
		\affiliation[inst1]{
			organization={School of Accounting and Auditing, Yunnan University of Finance and Economics},
			city={Kunming},
			postcode={650221},
			country={China}}
		
		\affiliation[inst2]{
			organization={School of Statistics and Mathematics, Yunnan University of Finance and Economics},
			city={Kunming},
			postcode={650221},
			country={China}}
		
		\affiliation[inst3]{
			organization={School of Engineering, Kunming University},
			city={Kunming},
			postcode={650214},
			country={China}}
		
		\affiliation[inst4]{
			organization={Institute of Technical Physics and Materials Science, Centre for Energy Research},
			city={Budapest},
			postcode={1525},
			country={Hungary}}
		
		\begin{abstract}
			The danger of collective risk in a public goods game could provide an effective escape route from the tragedy of the commons.
			Existing studies usually embed risk mitigation in the contribution itself, whereas many real systems rely on a separate costly and non-excludable protective activity that requires a participation quorum. Whether such protection can evolve and persist remains unclear. Here, we formulate an $N$-player public-goods game with ordinary cooperators, defectors, and protective cooperators. Defectors endogenously increase a saturating probability of collective failure, whereas protective cooperators incur an additional cost and reduce this risk for all group members once their number reaches the quorum. Replicator analysis and numerical bifurcation calculations yield three main findings. First, protective cooperation gains an advantage when the focal participant completes the quorum and the resulting expected loss reduction exceeds its additional cost. This advantage varies non-monotonically with group composition. Second, when a single protective cooperator cannot activate protection, protection effectiveness does not enter its first-order invasion fitness at full defection; greater effectiveness cannot overcome the rare-invasion barrier when full defection is locally asymptotically stable. Third, a boundary saddle--node bifurcation creates a saddle and a locally asymptotically stable coexistence equilibrium of defectors and protective cooperators. When full cooperation and full defection are also locally asymptotically stable, three attractors coexist, and the numerically observed evolutionary outcome depends on the initial population composition. These results distinguish the effectiveness of activated protection from its ability to become established from rarity.
		\end{abstract}
		
	\end{frontmatter}

	\section{Introduction}
	\label{sec:introduction}
	
	Cooperation allows individuals to produce collective benefits that are independently unattainable, yet it remains constantly threatened by free-riding. Explaining the emergence and maintenance of cooperation amid this tension is a fundamental pursuit in evolutionary theory~\citep{axelrod1981evolution,smith1982evolution}. To formally distill the essence of such social dilemmas, researchers 
	generally
	employ the public goods game (PGG). In this framework, cooperators pay a personal cost to generate a shared return, whereas defectors exploit this resource without bearing the associated costs~\citep{ szabo2007evolutionary, szolnoki2010reward}. Because the individual marginal return on any contribution is strictly lower than its initial cost, natural selection ultimately favors defection, driving the group toward a fundamentally worse outcome than if all had cooperated. Consequently, the PGG elegantly captures the evolutionary underpinnings of the tragedy of the commons~\citep{hardin1968tragedy,ostrom1990governing,perc_jrsi13}.
	
	Evolutionary game theory has identified several mechanisms that can offset the individual advantage of defection. Five canonical routes
	are kin selection~\citep{eberhard1975evolution}, direct reciprocity~\citep{trivers_qrb71}, indirect reciprocity~\citep{nowak2005evolution}, network reciprocity~\citep{nowak_n92b}, and group selection~\citep{wilson_ds_pnas75}. These mechanisms promote cooperation through distinct evolutionary channels. Kin selection relies on genetic relatedness, direct reciprocity on repeated interactions, indirect reciprocity on reputation, network reciprocity on structured populations, and group selection on differences in success among groups~\citep{nowak2006five,he_s_csf26,wang_cq_csf23,yue_h_csf25}. Incentive mechanisms provide an alternative 
	route by directly modifying the payoff difference between cooperation and defection. Costly punishment and reward can promote cooperation~\citep{fehr2000cooperation,boyd2003evolution,rand2009positive}, although their effectiveness may be limited by implementation costs and second-order free riding~\citep{hauert2007via,szolnoki2017second,panchanathan2004indirect}. To circumvent these implementation barriers, institutional and behavioral regulations -- including voluntary participation~\citep{hauert2002volunteering}, optional exit~\citep{shen2021exit,li2024granting}, and social exclusion~\citep{sasaki2013evolution,li2015social,li2024antisocial} -- offer alternative pathways by dynamically controlling entry into collective interactions or restricting access to shared returns. Collectively, these foundational studies demonstrate how assortment, incentives, and participation rules successfully tip the evolutionary balance in favor of cooperation within ordinary public goods games~\citep{perc2017statistical,wang_cq_c25}. However, a fundamentally distinct evolutionary challenge arises when defection not only diminishes the provision of a public good but simultaneously inflates the probability of a catastrophic collective failure that indiscriminately penalizes every group member.
	
	The collective-risk social dilemma formalizes this shared-loss structure by imposing a probabilistic loss when a group fails to reach a collective target~\citep{croson2000step,milinski2006stabilizing}. Experiments show that sufficiently severe risk can increase contributions, although high risk or scientific uncertainty alone does not guarantee collective success~\citep{milinski2008collective,barrett2012climate}. Evolutionary analyses further show that contribution targets create coordination barriers and that risk magnitude, group size, and target strictness can generate multiple evolutionary outcomes~\citep{pacheco2008evolutionary,wang2009emergence,santos2011risk,chen2012risk}. Subsequent extensions have examined inequality, intergenerational sustainability, communication, insurance, and feedback between population behavior and the risk environment~\citep{tavoni2011inequality,vasconcelos2014climate,hauser2014cooperating,zhang2015insurance,barfuss2020caring,liu2023coevolutionary,hua2024coevolutionary}. More recent studies have broadened this institutional perspective by introducing centralized monitoring and reporting, competing risk-sharing arrangements, adaptive collective targets, and joint feedback among cooperation, risk, and cost~\citep{he2019central,wang2025strategic,hua2026adaptive,xu2026adaptive,wang2026coevolutionary}. These studies demonstrate that institutional responses and endogenous feedback can reorganize the equilibrium structure and generate initial-condition-dependent outcomes. Nevertheless,
	typical
	public-good provision is generally not separated from a distinct, cost-bearing, and non-excludable protective activity whose effectiveness depends on its own participation quorum.
	
	In reality, averting catastrophic risk frequently demands independent specialized protective interventions that are structurally decoupled from standard cooperative contributions. Such dedicated mechanisms operate under distinct dynamics. They exact additional operating costs from participants and remain dormant until a strict participation quorum is satisfied. Once triggered, however, they generate a strictly non-excludable safety barrier that indiscriminately shields all group members, naturally including free riders. For example, building a cross-border epidemic surveillance network or public disaster-prevention infrastructure demands dedicated construction and maintenance expenditures, and such systems only operate effectively when the number of participating nodes reaches a predefined quorum. The evolutionary dynamics governing the emergence and stability of such a specialized protective mechanism -- featuring independent costs, threshold activation, and non-excludability -- remain a critical theoretical challenge
	in our understanding.
	
	To address this gap, we develop an $N$-player evolutionary public-goods game that explicitly separates risk mitigation from ordinary contributions. The population consists of ordinary cooperators, defectors, and protective cooperators. Both ordinary and protective cooperators contribute to the public good, while protective cooperators also incur an operating cost. The probability of collective failure increases and gradually saturates with the number of defectors, making risk endogenous to group composition. Protection reduces this probability for every group member, including defectors, but becomes active only when the number of protective cooperators reaches a prescribed threshold. We derive the expected payoffs under random matching, establish the local invasion and stability conditions of the three-strategy replicator system, and use numerical branch tracking to locate changes in its equilibrium structure. The results show that the advantage of protective cooperation varies non-monotonically with group composition because it is concentrated in configurations in which the focal individual completes the quorum. When a single protective cooperator cannot activate the mechanism, increasing protection effectiveness does not improve its first-order invasion fitness near full defection. Sufficiently effective protection can nevertheless create a locally asymptotically stable coexistence state at a positive protective-cooperator frequency through a boundary saddle--node bifurcation while the rare-invasion barrier remains intact. When full cooperation is also locally asymptotically stable, the resulting tristability makes the numerically observed outcome depend on the initial population composition.
	
	\section{Model and methods}
	\label{sec:model}
	
	\subsection{Population and strategies}
	\label{subsec:population-pgg}
	
	Consider an infinite, well-mixed population whose members repeatedly form groups of integer size $N\geq3$. Each individual adopts one of three available strategies, namely ordinary cooperation ($C$), defection ($D$), or protective cooperation ($S$). Denoting their respective population frequencies by $x$, $y$, and $z$, the evolutionary state space is formally defined as the simplex
	\[
	\Delta_2
	=
	\left\{
	(x,y,z)\in\mathbb R_{\geq0}^{3}
	\mid
	x+y+z=1
	\right\}.
	\]
	The well-mixed assumption naturally implies that the strategies of the other group members are sampled independently from these population frequencies.
	
	Within any realized group, individuals adopting either strategy $C$ or $S$ contribute a fixed cost $c>0$ to a public pool, whereas $D$-players contribute nothing. This aggregate contribution is multiplied by a synergy factor $r$ and distributed equally among all $N$ group members. By imposing the standard social-dilemma condition $1<r<N$, any individual contribution increases the collective payoff ($r>1$) yet yields a net personal deficit ($r/N<1$) in the hypothetical absence of collective risk. Furthermore, beyond the baseline contribution $c$, strategy $S$ incurs an independent operating cost $k>0$ per interaction, strictly regardless of whether the protective mechanism is ultimately triggered.
	
	\subsection{Collective risk and quorum-activated protection}
	\label{subsec:risk-quorum}
	
	Crucially, defection endogenously generates a shared risk of collective failure. For a realized group harboring $d$ defectors, this failure probability takes the explicit form
	\begin{equation}
		p(d)
		=
		1-e^{-\gamma d}
		=
		1-\rho^d,
		\qquad
		\gamma>0,
		\label{eq:risk}
	\end{equation}
	where $\gamma$ measures the sensitivity of collective failure risk to the number of defectors, and $\rho=e^{-\gamma}\in(0,1)$ is the multiplicative survival (no-failure) factor associated with one defector. Each additional defector multiplies the group's survival probability by $\rho$. Because the marginal increment $p(d+1)-p(d)=(1-\rho)\rho^d$ is positive and decreases with $d$, collective failure risk increases with the number of defectors at a diminishing rate.
	
	Should a collective failure occur, every group member suffers an identical additive loss $L>0$, a severe penalty that operates independently of the public-good returns and induces no wealth transfers between strategies. To mitigate this threat, collective protection becomes functionally active if and only if the number of protective cooperators meets a predefined integer quorum $M\in\{2,\ldots,N-1\}$. Upon activation, this mechanism attenuates the baseline failure probability by a protective efficacy fraction $\omega\in[0,1]$. Consequently, a group containing $d$ defectors faces a residual failure risk of $p(d)$ if the quorum remains unmet, and $(1-\omega)p(d)$ otherwise. Because this risk reduction blankets the entire group without excluding defectors, the activated protection acts as a pure public good rather than a targeted sanction.
	
	Restricting the quorum to the interior range $M\in\{2,\ldots,N-1\}$ deliberately isolates a nondegenerate coordination dilemma. Specifically, a minimal threshold of $M=1$ would allow a solitary mutant to activate protection, improperly embedding $\omega$ into the first-order invasion fitness near full defection. Conversely, demanding unanimous participation ($M=N$) mandates an all-$S$ group; such a composition natively lacks defectors and intrinsically faces a baseline risk of $p(0)=0$, thereby rendering any protective mechanism structurally redundant.
	
	\subsection{Payoffs in a realized group}
	\label{subsec:composition-payoffs}
	
	To formally characterize the interactions within a realized group, let $n=N-1$ denote the number of co-players interacting with a focal individual. We define $i$, $j$, and $m$ as the respective numbers of co-players adopting strategies $C$, $D$, and $S$. Consequently, this local configuration inherently satisfies the composition constraint $i+j+m=n$.
	
	The activation of the protective mechanism depends
	critically on whether the total number of protective cooperators within the group satisfies the predefined quorum threshold $M$. Specifically, for a focal $C$- or $D$-player, protection is triggered if and only if the number of $S$-players among its co-players reaches this quorum (i.e., $m \geq M$). Conversely, because a focal $S$-player actively contributes to the protective mechanism, this focal individual naturally counts toward the quorum, thereby relaxing the required threshold among its co-players to $m \geq M-1$.
	
	Letting $\mathbb{I}_{\mathcal{E}}$ denote the standard indicator function -- yielding $1$ if event $\mathcal{E}$ occurs and $0$ otherwise -- the expected payoffs for the three strategies, conditional on the realized co-player composition, are formulated as
	\begin{equation}
		\left\{
		\begin{aligned}
			\pi_C&=\frac{rc(i+m+1)}{N}-c
			-Lp(j)\left[1-\omega\mathbb{I}_{m\geq M}\right],
			\\
			\pi_D&=\frac{rc(i+m)}{N}
			-Lp(j+1)\left[1-\omega\mathbb{I}_{m\geq M}\right],
			\\
			\pi_S&=\frac{rc(i+m+1)}{N}-c-k
			-Lp(j)\left[1-\omega\mathbb{I}_{m\geq M-1}\right].
		\end{aligned}
		\right.
		\label{eq:single-payoffs}
	\end{equation}
	
	Structurally, the first term in each payoff expression represents the focal player's proportionate share of the generated public good. Regarding individual expenditures, both $C$- and $S$-players incur the baseline contribution cost $c$, whereas the focal $S$-player bears an additional operational expense $k$ to maintain the protection. Furthermore, because a focal defector intrinsically raises the total count of free riders in the group from $j$ to $j+1$, the corresponding collective failure probability evaluated in $\pi_D$ escalates to $p(j+1)$. Finally, the concluding bracketed term quantifies the expected collective loss, utilizing the indicator functions to precisely encode the distinct, focal-strategy-specific quorum conditions required to activate the protective mechanism.
	\subsection{Expected payoffs under random matching}
	\label{subsec:expected-payoffs}
	
	Assuming a well-mixed population where individuals interact through random matching, the co-player composition confronting a focal individual follows a multinomial distribution given by $(i,j,m) \sim \operatorname{Multinomial}(n;x,y,z)$. Accordingly, the expected population-level payoff for any focal strategy $\ell\in\{C,D,S\}$ is defined as $P_\ell = \mathbb E\!\left[\pi_\ell(i,j,m)\right]$. The exact analytical evaluations of these multinomial expectations are systematically provided in~\ref{app:payoffs}.
	
	To facilitate the calculation of collective failure risks, we define an auxiliary variable representing the expected non-failure factor $q = x+z+\rho y = 1-(1-\rho)y$. Because any randomly sampled co-player contributes a multiplicative non-failure factor of $1$ (if adopting $C$ or $S$) or $\rho$ (if adopting $D$), its expected value is precisely $q$. Given that the $n$ co-players are drawn independently, we obtain the fundamental expectation relation $q^n = \mathbb E[\rho^j] = \mathbb E[e^{-\gamma j}]$, where the number of defectors naturally follows a binomial distribution $j\sim\operatorname{Binomial}(n,y)$.
	
	Furthermore, to capture the threshold-dependent activation of the protective mechanism, we introduce the risk-weighted activation terms for a generic integer threshold $h\in\{0,\ldots,n\}$ as follows:
	\begin{equation}
		T_h
		=
		\mathbb E\!\left[
		p(j)\mathbb I_{m\geq h}
		\right],
		\qquad
		T_h^D
		=
		\mathbb E\!\left[
		p(j+1)\mathbb I_{m\geq h}
		\right].
		\label{eq:quorum-risk-expectations}
	\end{equation}
	Crucially, these formulations represent risk-weighted activation terms rather than mere activation probabilities, as each threshold-satisfying composition is explicitly weighted by its corresponding state-dependent failure probability. Consequently, while $\omega T_h$ and $\omega T_h^D$ capture the expected reductions in the collective failure probability, $L\omega T_h$ and $L\omega T_h^D$ strictly quantify the actual expected losses successfully averted by the protection. The exact finite-sum expressions for these terms are derived in~\ref{app:quorum-terms}.
	
	By setting $h=0$, we effectively remove the activation filter, which yields:
	\begin{equation}
		T_0
		=
		1-q^n,
		\qquad
		T_0^D
		=
		1-\rho q^n.
		\label{eq:T0}
	\end{equation}
	In this context, $T_0$ signifies the expected baseline unprotected failure probability faced by a focal contributor, whereas $T_0^D$ incorporates the marginal risk increment induced by a focal defector.
	
	Aggregating these mathematical components, the expected population-level payoffs for the three strategies are formulated as:
	\begin{equation}
		\left\{
		\begin{aligned}
			P_C
			&=
			\frac{rc[1+n(x+z)]}{N}-c
			-L\left[T_0-\omega T_M\right],\\
			P_D
			&=
			\frac{rcn(x+z)}{N}
			-L\left[T_0^D-\omega T_M^D\right],\\
			P_S
			&=
			\frac{rc[1+n(x+z)]}{N}-c-k
			-L\left[T_0-\omega T_{M-1}\right].
		\end{aligned}
		\right.
		\label{eq:PS}
	\end{equation}
	
	Structurally, the first component in each payoff equation captures the expected share of the public-good return, the second component accounts for the strategy-specific investment costs, and the final bracketed term reflects the residual expected collective loss after accounting for the protective mechanism. Notably, the activation index shifts from $M$ to $M-1$ exclusively for a focal protective cooperator ($P_S$), accurately reflecting the endogenous property that the focal individual's own participation counts toward satisfying the critical activation quorum.
	
	\subsection{Selection gradients}
	\label{subsec:selection-gradients}
	
	To analytically isolate the marginal protection benefit uniquely conferred by strategy $S$, we define the pivotal threshold probability as $\Psi_M := T_{M-1}-T_M$. Because the events $m\geq M-1$ and $m\geq M$ differ exclusively at the discrete boundary $m=M-1$, this quantity can be explicitly expanded as
	\begin{equation}
		\begin{aligned}
			\Psi_M
			&=
			\mathbb E\!\left[
			p(j)\mathbb I_{m=M-1}
			\right]
			\\
			&=
			\binom{n}{M-1}z^{M-1}
			\left[
			(1-z)^{n-M+1}
			-(x+\rho y)^{n-M+1}
			\right].
		\end{aligned}
		\label{eq:pivotal-expectation}
	\end{equation}
	The quantity $\Psi_M$ represents the risk-weighted probability that the focal protective cooperator completes the activation threshold. Consequently, the payoff difference between protective and ordinary cooperators reduces to the exact identity
	\begin{equation}
		P_S-P_C
		=
		-k+L\omega\Psi_M.
		\label{eq:pivotal-identity}
	\end{equation}
	Here, $L\omega\Psi_M$ is the expected collective loss avoided because the focal player is pivotal for activation. Protective cooperation is favored over ordinary cooperation if and only if this expected pivotal benefit exceeds the fixed operating cost $k$.
	
	To facilitate the subsequent dynamical analysis on the state space, we eliminate the frequency of defectors by substituting $y=1-x-z$. From this point onward, the variables $q$, $R_M$, and $\Psi_M$ denote their corresponding composite functions evaluated on the reduced two-dimensional simplex, ensuring that all partial derivatives with respect to $x$ and $z$ are strictly taken within this constrained domain. We then define the selection gradients relative to defection as
	\begin{equation}
		A(x,z;\omega)
		=
		P_C-P_D,
		\qquad
		B(x,z;\omega)
		=
		P_S-P_D.
		\label{eq:payoff-differences}
	\end{equation}
	For notational brevity, all structural parameters other than the protection effectiveness $\omega$ are suppressed from the function arguments.
	
	To express these selection gradients in closed form, we introduce the baseline marginal payoff parameter $a = \frac{rc}{N}-c$ alongside the threshold-dependent expectation term $R_M = \mathbb E\!\left[\rho^j\mathbb I_{m\geq M}\right]$. As formally derived in~\ref{app:payoff-differences}, substituting these components yields the exact closed-form expressions:
	\begin{align}
		A(x,z;\omega)
		&=
		a+L(1-\rho)
		\left[
		q^n-\omega R_M
		\right],
		\label{eq:A}\\
		B(x,z;\omega)
		&=
		A(x,z;\omega)-k+L\omega\Psi_M.
		\label{eq:B}
	\end{align}
	The constant $a$ captures the direct payoff consequence of replacing a defector with a contributor in the absence of collective risk. The bracketed term in $A(x,z;\omega)$ gives the associated reduction in expected collective loss, including groups in which protection is already active. The function $B(x,z;\omega)$ additionally incorporates the operating cost $k$ and the risk-weighted pivotal benefit $L\omega\Psi_M$ specific to protective cooperation.
	
	\subsection{Replicator dynamics}
	\label{subsec:replicator}
	
	To model the evolutionary trajectories of the competing strategies, we first establish the mean population payoff, defined as $\bar P = xP_C+yP_D+zP_S$. By absorbing an arbitrary positive selection rate into the time scale without loss of generality, the temporal evolution of the strategy frequencies is governed by the standard replicator equations~\citep{taylor1978evolutionary,hofbauer1998evolutionary}:
	\begin{equation}
		\left\{
		\begin{aligned}
			\dot x&=x(P_C-\bar P),\\
			\dot y&=y(P_D-\bar P),\\
			\dot z&=z(P_S-\bar P),
		\end{aligned}
		\right.
		\label{eq:replicator-three}
	\end{equation}
	where the overdot denotes differentiation with respect to evolutionary time.
	
	Because the population frequencies must rigorously sum to unity, eliminating the defector frequency via $y=1-x-z$ allows us to mathematically project the dynamics onto the reduced two-dimensional state space, formally defined as the simplex $\Delta_{xz} = \left\{ (x,z)\in\mathbb R_{\geq0}^{2} \mid x+z\leq1 \right\}$. By substituting the previously derived selection gradients from Eq.~\eqref{eq:payoff-differences} into this framework, the continuous-time evolution on the reduced simplex can be explicitly expressed as the planar system:
	\begin{equation}
		\left\{
		\begin{aligned}
			\dot x
			&=
			x\left[
			(1-x)A(x,z;\omega)-zB(x,z;\omega)
			\right],
			\\
			\dot z
			&=
			z\left[
			(1-z)B(x,z;\omega)-xA(x,z;\omega)
			\right].
		\end{aligned}
		\right.
		\label{eq:planar}
	\end{equation}
	
	Structurally, the boundaries of this simplex constitute invariant manifolds. Specifically, the coordinate edges $x=0$ and $z=0$ remain strictly invariant because the evolutionary growth rate of any strategy is inherently scaled by its own frequency, thereby preventing an absent strategy from spontaneously emerging. Similarly, the diagonal boundary spanning $x+z=1$ is dynamically invariant, as this condition corresponds to a population entirely devoid of defectors ($y=0$), rendering $\dot y=0$ within the original full system in Eq.~\eqref{eq:replicator-three}. Together, Equations~\eqref{eq:A}, \eqref{eq:B}, and \eqref{eq:planar} form the closed, rigorously defined planar dynamical system that serves as the analytical foundation for all subsequent stability and bifurcation analyses.
	
	\subsection{Equilibria and local stability}
	\label{subsec:equilibria-stability}
	
	To facilitate a rigorous stability analysis of the dynamical system, we first rewrite the planar equations \eqref{eq:planar} in the generalized vector field form $\dot x=F(x,z;\omega)$ and $\dot z=G(x,z;\omega)$, and correspondingly define the associated Jacobian matrix of the system as
	\begin{equation}
		J(x,z;\omega)
		=
		\begin{pmatrix}
			F_x&F_z\\
			G_x&G_z
		\end{pmatrix}.
		\label{eq:J-main}
	\end{equation}
	The explicit analytical expressions for the general Jacobian, alongside its specific reductions evaluated at both boundary and interior equilibria, are systematically derived in~\ref{app:jacobian}.
	
	Following standard nonlinear dynamical systems theory, an equilibrium point is classified as locally asymptotically stable if and only if both eigenvalues of its evaluated Jacobian matrix possess strictly negative real parts. Conversely, the presence of at least one eigenvalue with a positive real part renders the equilibrium unstable. In non-hyperbolic scenarios where the leading eigenvalue possesses a zero real part, first-order linearization becomes structurally inconclusive, thereby necessitating the examination of higher-order terms or center manifold reductions to rigorously ascertain the system's local topological behavior.
	
	The homogeneous populations consisting entirely of defectors, ordinary cooperators, and protective cooperators are denoted by $\mathbf{D}=(0,1,0)$, $\mathbf{C}=(1,0,0)$, and $\mathbf{S}=(0,0,1)$. They correspond to the vertices $(0,0)$, $(1,0)$, and $(0,1)$ of the reduced simplex, respectively.
	
	\subsubsection{Homogeneous equilibria}
	\label{subsubsec:homogeneous-equilibria}
	
	At the full-defection equilibrium $\mathbf{D}$, we define the baseline boundary selection gradient as $A_D := A(0,0;\omega) = a+L(1-\rho)\rho^n$. Evaluating the Jacobian at this state yields the two transversal invasion eigenvalues:
	\begin{equation}
		\lambda_C^D=A_D,
		\qquad
		\lambda_S^D=A_D-k.
		\label{eq:D-eigen}
	\end{equation}
	The first eigenvalue gives the initial growth rate of a rare ordinary cooperator in an otherwise defecting population, whereas the second gives the corresponding growth rate of a rare protective cooperator. Consequently, full defection is locally asymptotically stable if and only if $A_D<0$, is a saddle when $0<A_D<k$, and is an unstable node when $A_D>k$. The boundary equality cases are nonhyperbolic and are analyzed in~\ref{app:nonhyperbolic-vertices}.
	
	Notably, neither eigenvalue in Eq.~\eqref{eq:D-eigen} incorporates the protection effectiveness parameter $\omega$. Because the activation quorum satisfies $M\geq2$, a solitary mutant protective cooperator is inherently incapable of triggering the collective protective mechanism. Therefore, amplifying the protection effectiveness cannot alter the first-order invasion fitness at full defection, preserving a strict rare-invasion barrier.
	
	Similarly, at the full-cooperation equilibrium $\mathbf{C}$, we define the analogous boundary gradient $A_C := A(1,0;\omega) = a+L(1-\rho)$. The corresponding invasion eigenvalues governing the introduction of alternative strategies are determined as
	\begin{equation}
		\lambda_D^C=-A_C, \qquad \lambda_S^C=-k.
		\label{eq:C-eigenvalues}
	\end{equation}
	Thus, the full-cooperation state achieves local asymptotic stability precisely when $A_C>0$, and it degenerates into a saddle point when $A_C<0$. In the critical scenario where $A_C=0$, the state becomes nonhyperbolic;~\ref{app:nonhyperbolic-vertices} formally establishes its instability by analyzing the flow dynamics strictly along the invariant $C$--$D$ edge.
	
	Finally, at the full-protective-cooperation equilibrium $\mathbf{S}$, the exact invasion eigenvalues are determined as
	\begin{equation}
		\lambda_C^S = k, \qquad \lambda_D^S = k-a-L(1-\rho)(1-\omega).
		\label{eq:S-eigenvalues}
	\end{equation}
	Because the invasion fitness of an ordinary cooperator is strictly positive ($\lambda_C^S=k>0$ for any maintained operating cost $k>0$), the full-protective-cooperation state is unconditionally unstable across all permissible parameter regimes. Specifically, the system exhibits a saddle point if the secondary eigenvalue is negative ($\lambda_D^S<0$) and an unstable node if it is positive ($\lambda_D^S>0$). When $\lambda_D^S=0$, the equilibrium remains globally unstable albeit passing through a nonhyperbolic topological threshold.
	
	\subsubsection{Boundary equilibria}
	\label{subsubsec:boundary-equilibria}
	
	Restricting the evolutionary dynamics to the invariant $C$--$D$ edge where $z=0$, the differential equation governing the fraction of ordinary cooperators simplifies to $\dot x = x(1-x)A(x,0;\omega)$. As shown in~\ref{app:boundary-stability}, the restricted gradient $A(x,0;\omega)$ is strictly increasing in $x$. Consequently, a unique nontrivial edge equilibrium, denoted by $E_{CD} = (x_{CD},1-x_{CD},0)$, exists if and only if $A_D<0<A_C$. Its tangential eigenvalue is positive, whereas its transverse protective-cooperator invasion eigenvalue is $-k<0$; hence, $E_{CD}$ is a saddle. Along the invariant edge, this saddle separates the one-dimensional basins of full cooperation and full defection.
	
	Turning to the invariant $C$--$S$ edge, the complete absence of defectors implies that the collective failure probability functionally vanishes ($p(0)=0$), which correspondingly nullifies the pivotal threshold probability ($\Psi_M=0$). Under these ideal conditions, the payoff differential identity in Eq.~\eqref{eq:pivotal-identity} tightly reduces to $P_S-P_C=-k<0$. This structural asymmetry guarantees that no nontrivial $C$--$S$ equilibrium can exist for any strictly positive operating cost ($k>0$); thus, the evolutionary flow along this boundary is unidirectionally driven from $S$ toward $C$. The degenerate, cost-free scenario where $k=0$ entails structural sensitivities and is separately addressed in~\ref{app:zero-cost}.
	
	Along the invariant $D$--$S$ edge, any nontrivial coexistence equilibrium must take the generic form $E_{DS} = (0,1-z_{DS},z_{DS})$ for $0<z_{DS}<1$, and must strictly satisfy the root condition $B(0,z_{DS};\omega)=0$. By linearizing the system around this boundary state, the transverse and tangential eigenvalues are respectively given by
	\begin{equation}
		\lambda_{\perp} = A(0,z_{DS};\omega), \qquad \lambda_{\parallel} = z_{DS}(1-z_{DS})B_z(0,z_{DS};\omega).
		\label{eq:DS-eigenvalues}
	\end{equation}
	The transverse eigenvalue $\lambda_{\perp}$ gives the invasion growth rate of a rare ordinary cooperator, whereas the tangential eigenvalue $\lambda_{\parallel}$ governs perturbations along the invariant edge. Assuming hyperbolicity, the $D$--$S$ equilibrium is locally asymptotically stable if and only if $A(0,z_{DS};\omega)<0$ and $B_z(0,z_{DS};\omega)<0$. It is a saddle when the two eigenvalues have opposite signs, while the vanishing of either eigenvalue makes first-order linearization inconclusive.
	
	A critical topological transition, characterized as a boundary saddle--node bifurcation, unfolds when two distinct nontrivial $D$--$S$ equilibria mutually approach and coalesce. Analytically, this bifurcation is pinpointed by the simultaneous satisfaction of the equilibrium and tangency conditions:
	\begin{equation}
		B(0,z_{DS};\omega)=0,
		\qquad
		B_z(0,z_{DS};\omega)=0.
		\label{eq:DSfold}
	\end{equation}
	To ensure that the tangential zero eigenvalue remains geometrically simple within the full two-dimensional planar system, the transverse non-degeneracy condition must hold:
	\begin{equation}
		A(0,z_{DS};\omega)\neq0.
		\label{eq:DSfold-transverse}
	\end{equation}
	Furthermore, treating the protection effectiveness $\omega$ as the primary bifurcation parameter while holding all other parameters fixed, the definitive existence of a nondegenerate codimension-one fold strictly imposes two additional transversality requirements:
	\begin{equation}
		B_{zz}(0,z_{DS};\omega)\neq0,
		\qquad
		B_\omega(0,z_{DS};\omega)\neq0.
		\label{eq:DSfold-nondegeneracy}
	\end{equation}
	The rigorous derivation of these bifurcation criteria, grounded in the scalar vector field properly restricted to the $D$--$S$ edge, is systematically detailed in~\ref{app:boundary-fold-reduction}.
	
	\subsubsection{Interior equilibria}
	\label{subsubsec:interior-equilibria}
	
	By definition, a fully polymorphic (interior) equilibrium, denoted as $E_*=(x^*,y^*,z^*)$, requires the strictly positive coexistence of all three strategies ($x^*,y^*,z^*>0$). Within the replicator dynamics framework, this mutual coexistence necessitates the exact equalization of all three expected payoffs, which is algebraically equivalent to the simultaneous vanishing of the respective selection gradients:
	\begin{equation}
		A(x^*,z^*;\omega)=0,
		\qquad
		B(x^*,z^*;\omega)=0.
		\label{eq:interior-conditions}
	\end{equation}
	As systematically detailed in~\ref{app:interior-jacobian}, the Jacobian matrix evaluated at such an interior state admits a mathematically tractable factorization. Specifically, it can be analytically shown that the determinant of this evaluated Jacobian, $\det J_*$, simplifies to the exact form
	\begin{equation}
		\det J_*
		=
		x^*y^*z^*
		\left.
		\left(
		A_xB_z-A_zB_x
		\right)
		\right|_{E_*}.
		\label{eq:Jdet}
	\end{equation}
	
	Applying standard linear stability criteria, a hyperbolic interior equilibrium achieves local asymptotic stability if and only if its Jacobian simultaneously possesses a positive determinant ($\det J_*>0$) and a negative trace ($\operatorname{tr}J_*<0$). Conversely, the state structurally manifests as a saddle point whenever its determinant is strictly negative ($\det J_*<0$), and acts as a locally unstable source (or unstable spiral) if both the determinant and the trace are positive ($\det J_*>0$ and $\operatorname{tr}J_*>0$). In the critical event of a singular Jacobian matrix ($\det J_*=0$), or a pure imaginary pair of eigenvalues ($\operatorname{tr}J_*=0$ with $\det J_*>0$), the equilibrium degenerates into a nonhyperbolic state, rendering first-order linearization dynamically inconclusive.
	
	A prospective topological transition involving an interior saddle--node bifurcation is mathematically pinpointed when an equilibrium candidate simultaneously satisfies the payoff equality in Eq.~\eqref{eq:interior-conditions} and exhibits a singular Jacobian matrix. Because the strategy frequencies $x^*, y^*, z^*$ are strictly positive in the interior domain, this singularity condition is algebraically equivalent to the vanishing of the determinant of the underlying selection gradients:
	\begin{equation}
		\left.
		\left(
		A_xB_z-A_zB_x
		\right)
		\right|_{E_*}
		=
		0.
		\label{eq:interior-fold}
	\end{equation}
	To ensure this zero eigenvalue remains geometrically simple within the planar system, the trace of the Jacobian must concurrently remain non-zero:
	\begin{equation}
		\operatorname{tr}J_*\neq0.
		\label{eq:interior-fold-simple}
	\end{equation}
	While these specific criteria successfully isolate an equilibrium with a simple zero eigenvalue, they are inherently insufficient to definitively prove the occurrence of a nondegenerate saddle--node bifurcation. To rigorously confirm this structural transition, the requisite parameter transversality and quadratic nondegeneracy conditions are independently formulated and computationally verified in~\ref{app:fold-detection}.
	
	Finally, Table~\ref{tab:equilibrium-summary} systematically catalogues the diverse equilibrium classes subsequently employed in the numerical bifurcation analysis. Unless explicitly stated otherwise, it should be understood that all stability classifications provided herein strictly pertain to local topological properties within the reduced two-dimensional strategy simplex.
	\begin{table*}[t]
		\centering
		\small
		\renewcommand{\arraystretch}{1.4} 
		\caption{Existence conditions and local classification of the equilibrium classes on the strategy simplex.}
		\label{tab:equilibrium-summary}
		
		\begin{tabular}{
				>{\raggedright\arraybackslash}p{0.18\textwidth}
				>{\raggedright\arraybackslash}p{0.26\textwidth}
				>{\raggedright\arraybackslash}p{0.48\textwidth}
			}
			\toprule
			\textbf{Equilibrium class} & 
			\textbf{Existence condition} & 
			\textbf{Local classification within the simplex} \\
			\midrule
			
			$\mathbf{D} = (0,1,0)$ &
			All admissible parameters &
			Locally asymptotically stable if $A_D < 0$; a saddle if $0 < A_D < k$; an unstable node if $A_D > k$; nonhyperbolic and unstable if $A_D = 0$ or $A_D = k$. \\
			\addlinespace 
			
			$\mathbf{C} = (1,0,0)$ &
			All admissible parameters &
			Locally asymptotically stable if $A_C > 0$; a saddle if $A_C < 0$; nonhyperbolic and unstable if $A_C = 0$. \\
			\addlinespace
			
			$\mathbf{S} = (0,0,1)$ &
			All admissible parameters &
			Unstable because $\lambda_C^S = k > 0$; a saddle if $\lambda_D^S < 0$; an unstable node if $\lambda_D^S > 0$; nonhyperbolic and unstable if $\lambda_D^S = 0$. \\
			\addlinespace
			
			$E_{CD}$ &
			$A_D < 0 < A_C$ &
			A saddle for $k > 0$, with one positive tangential eigenvalue and the negative transverse eigenvalue $-k$. \\
			\addlinespace
			
			Nontrivial $C$--$S$ equilibrium &
			None for $k > 0$; an equilibrium continuum exists for $k = 0$. &
			For $k > 0$, the edge flow is directed from $S$ toward $C$. For $k = 0$, the edge is neutral in the tangential direction and its transverse stability is parameter dependent. \\
			\addlinespace

			$E_{DS} = (0,1-z_{DS},z_{DS})$ &
			$B(0,z_{DS};\omega)=0$, with $0<z_{DS}<1$ &
			The transverse and tangential eigenvalues are
			$\lambda_{\perp}=A(0,z_{DS};\omega)$ and
			$\lambda_{\parallel}
			=z_{DS}(1-z_{DS})B_z(0,z_{DS};\omega)$, respectively.
			The equilibrium is locally asymptotically stable if both eigenvalues
			are negative, an unstable node if both are positive, and a saddle if
			they have opposite signs. It is nonhyperbolic if either eigenvalue
			vanishes. \\
			\addlinespace
			
			$E_* = (x^*,y^*,z^*)$ &
			$A(x^*,z^*;\omega) = B(x^*,z^*;\omega) = 0$, with $x^*, y^*, z^* > 0$. &
			For a hyperbolic equilibrium, locally asymptotically stable if $\operatorname{tr}J_* < 0$ and $\det J_* > 0$; locally unstable if $\operatorname{tr}J_* > 0$ and $\det J_* > 0$; and a saddle if $\det J_* < 0$. The equilibrium is nonhyperbolic if $\det J_* = 0$, or if $\operatorname{tr}J_* = 0$ with $\det J_* > 0$. \\
			
			\bottomrule
		\end{tabular}
	\end{table*}
	
	\section{Numerical bifurcation results}
	\label{sec:numerics}
	
	Unless stated otherwise, the numerical branch tracking and phase-space calculations use the baseline parameter set
	\begin{equation}
		\begin{aligned}
			N &= 5,\quad M = 2,\quad r = 2.3,\quad c = 1, \\
			k &= 0.4,\quad L = 4,\quad \gamma = 1.4.
		\end{aligned}
		\label{eq:baseline}
	\end{equation}
	
	\subsection{Two saddle--node bifurcations}
	\label{subsec:two-folds}
	
	\begin{figure*}[t]
		\centering
		\includegraphics[width=0.86\linewidth]{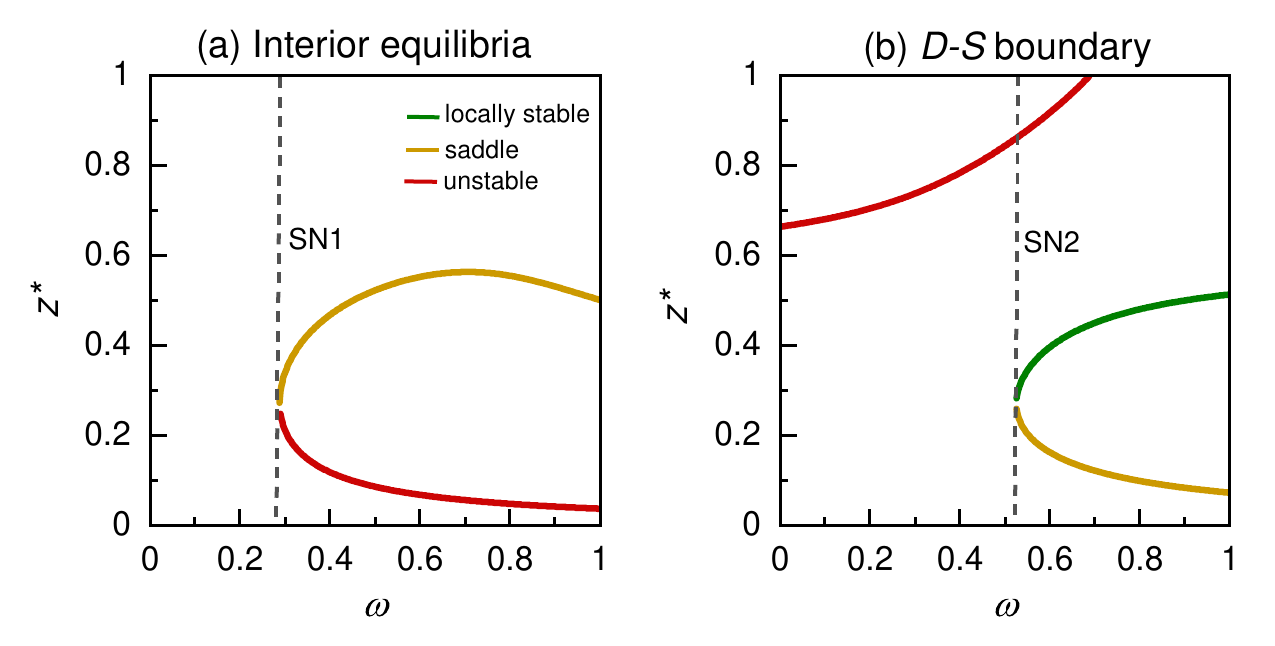}
		\caption{
			\textbf{Only the boundary saddle--node creates a locally stable protective
				equilibrium at positive protective-cooperator frequency, whereas the interior
				saddle--node adds no attractor.}
			Equilibrium branches with respect to protection effectiveness $\omega$
			under the baseline parameters in Eq.~\eqref{eq:baseline}.
			Panel~(a) shows the protective-cooperator frequency $z^*$ of the interior
			equilibria, and panel~(b) shows equilibria on the invariant
			$D$--$S$ edge. Green, orange, and red curves denote locally
			asymptotically stable, saddle, and unstable branches, respectively.
			The dashed vertical lines mark the interior saddle--node SN1 at
			$\omega_{\mathrm{SN1}}\approx0.2882$ and the boundary saddle--node
			SN2 at $\omega_{\mathrm{SN2}}\approx0.5254$. The unstable
			$D$--$S$ branch reaches an edge transcritical bifurcation at $\omega\approx0.6881$. The pure-$C$ and pure-$D$ equilibria are
			omitted because their locations and local classifications do not
			change with $\omega$ under the baseline parameters.
		}
		\label{fig:continuation}
	\end{figure*}
	
	Figure~\ref{fig:continuation} shows how the equilibrium set changes
	with $\omega$. Panel~(a) reports the interior equilibria, and
	panel~(b) reports equilibria on the invariant $D$--$S$ edge
	($x^*=0$ and $y^*=1-z^*$). The pure $C$ and $D$ vertices are omitted
	because their locations are independent of $\omega$ and both remain
	locally asymptotically stable under the baseline parameters. The
	branch search, stability classification, and fold tests are described
	in~\ref{app:numerical-continuation}.
	
	The first bifurcation, SN1, is a codimension-one saddle--node
	bifurcation in the interior of the simplex at
	\begin{equation*}
		(x^*,y^*,z^*,\omega_{\mathrm{SN1}})
		=
		(0.2994,0.4272,0.2734,0.2882).
	\end{equation*}
	No interior equilibrium was detected below $\omega_{\mathrm{SN1}}$
	within the continuation interval.
	Crossing SN1 creates a saddle and an unstable node. At the fold, the
	Jacobian has a simple zero eigenvalue and a positive second
	eigenvalue, while the numerical normal-form coefficients are nonzero.
	Neither branch is locally attracting. SN1 therefore changes the
	stable-manifold arrangement visible in the phase portraits without
	increasing the number of local attractors.
	
	The second bifurcation, SN2, occurs on the invariant $D$--$S$ edge at
	\begin{equation*}
		(x^*,y^*,z^*,\omega_{\mathrm{SN2}})
		=
		(0,0.7282,0.2718,0.5254).
	\end{equation*}
	At this point, the boundary equilibrium and tangency conditions hold:
	\begin{equation*}
		B(0,z^*;\omega_{\mathrm{SN2}})=0,
		\qquad
		B_z(0,z^*;\omega_{\mathrm{SN2}})=0.
	\end{equation*}
	The derivative tests in~\ref{app:fold-detection} confirm
	that SN2 is nondegenerate. For $\omega>\omega_{\mathrm{SN2}}$, the
	fold produces two $D$--$S$ equilibria. The lower-$z$ branch is a
	saddle. On the upper-$z$ branch, both the transverse eigenvalue
	$\lambda_{\perp}=A(0,z^*;\omega)$ and the tangential eigenvalue
	$\lambda_{\parallel}=z^*(1-z^*)B_z(0,z^*;\omega)$ are negative, so
	this branch is locally asymptotically stable within the simplex.
	Because the pure $C$ and $D$ vertices remain locally stable, the
	baseline system changes from bistability to tristability after SN2.
	A separate unstable $D$--$S$ branch reaches the pure-$S$ vertex at
	$\omega\approx0.6881$. Because this vertex collision does not change
	the number of local attractors, it is not involved in the formation
	of the protective state.
	\subsection{Phase-space reorganization and tristability}
	
	\begin{figure*}[t]
		\centering
		\includegraphics[width=.96\textwidth]{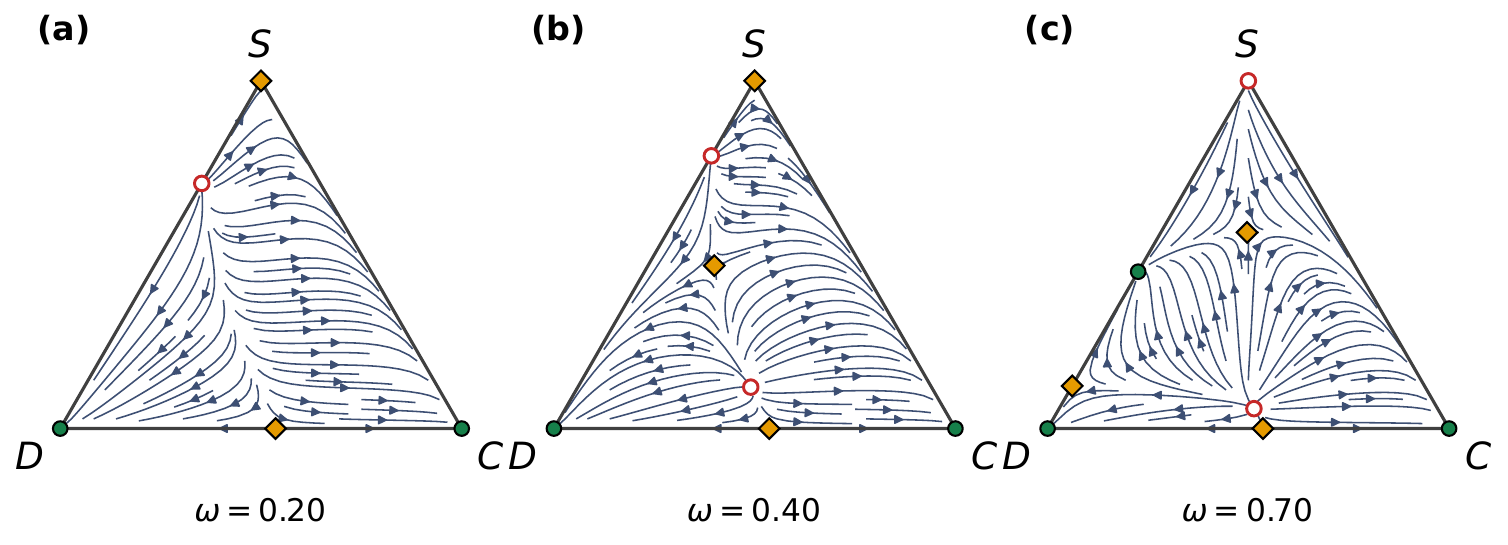}
		\caption{
			\textbf{The interior saddle--node reorganizes the separatrix geometry without adding an attractor, whereas the boundary saddle--node changes bistability into tristability by creating a stable $D$--$S$ state.}
			Replicator phase portraits on the equilateral strategy simplex under
			the baseline parameters at $\omega=0.20$ in panel~(a),
			$\omega=0.40$ in panel~(b), and $\omega=0.70$ in panel~(c).
			The vertices $C$, $D$, and $S$ represent the three homogeneous
			states, and arrowed integral curves indicate forward-time flow.
			Filled green circles, orange diamonds, and open red circles denote
			locally asymptotically stable equilibria, saddles, and unstable
			equilibria, respectively.
		}
		\label{fig:phase}
	\end{figure*}
	
	Figure~\ref{fig:phase} shows the corresponding changes in the simplex
	flow. At $\omega=0.20$ [panel~(a)], $C$ and $D$ are locally
	asymptotically stable, whereas $S$ is a saddle with eigenvalues
	$-1.4709$ and $0.4000$. The observed trajectories approach either
	$C$ or $D$, with the stable manifold of the $C$--$D$ edge saddle
	separating their basins. On the defector-free $C$--$S$ edge,
	Eq.~\eqref{eq:pivotal-identity} reduces to $P_S-P_C=-k<0$, so the edge
	flow points from $S$ toward $C$.
	
	At $\omega=0.40$ [panel~(b)], SN1 has created the interior saddle and
	unstable node, while SN2 has not yet occurred. The stable manifold of
	the interior saddle changes the numerically observed separatrix
	geometry, but the only local attractors remain $C$ and $D$.
	
	At $\omega=0.70$ [panel~(c)], after SN2, the locally asymptotically
	stable upper $D$--$S$ equilibrium is
	\begin{equation*}
		E_{DS}(0.70)
		=
		(0,0.5488,0.4512).
	\end{equation*}
	Equation~\eqref{eq:DS-eigenvalues} gives
	$\lambda_{\perp}=-0.4229$ and $\lambda_{\parallel}=-0.5349$.
	Both are real and negative, so $E_{DS}$ is a locally asymptotically
	stable node and the linearized relaxation near this equilibrium is
	non-oscillatory.
	
	\subsection{Parameter dependence of locally stable boundary protection}
	\label{subsec:effectiveness-threshold}
	\begin{figure*}[t]
		\centering
		\includegraphics[width=\textwidth]
		{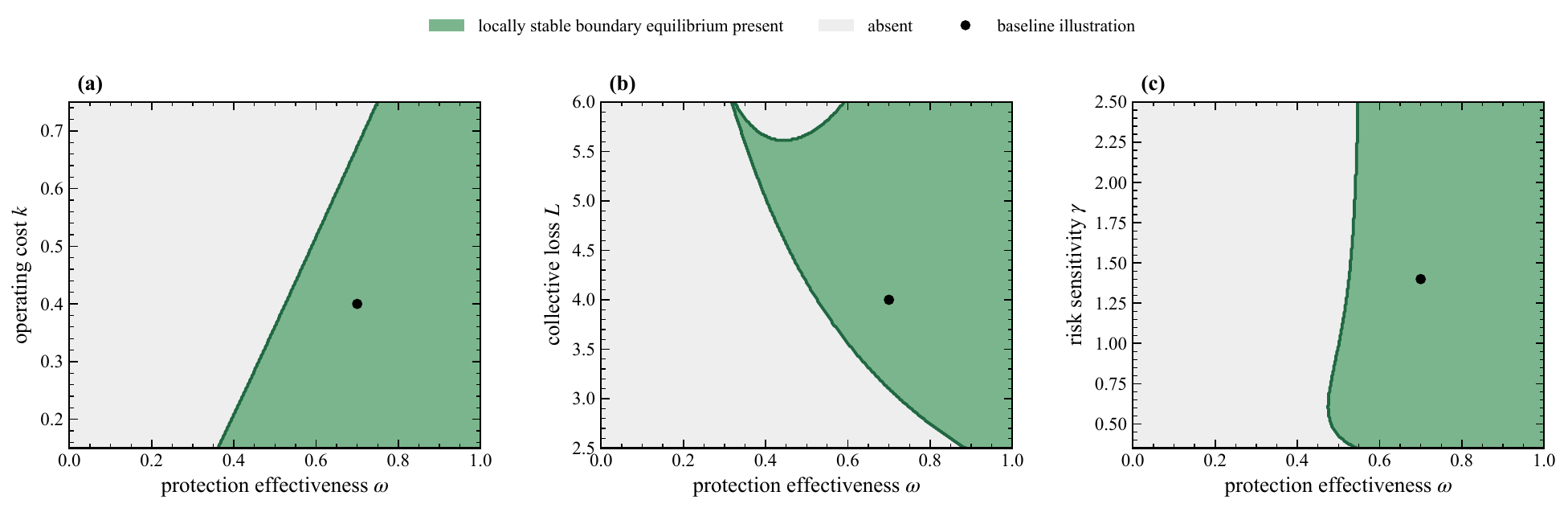}
		\caption{\textbf{Pointwise classification of locally asymptotically stable,	nontrivial $D$--$S$ equilibria in three two-parameter planes.}Panel~(a) shows the $(\omega,k)$ plane at $L=4$ and
			$\gamma=1.4$, panel~(b) shows the $(\omega,L)$ plane at
			$k=0.4$ and $\gamma=1.4$, and panel~(c) shows the
			$(\omega,\gamma)$ plane at $k=0.4$ and $L=4$.
			Green points contain at least one boundary root satisfying
			$\lambda_{\perp}<0$ and $\lambda_{\parallel}<0$, whereas no
			such root is detected at gray points. Solid contours are
			numerically extracted boundaries between these two
			classifications and need not represent a single
			saddle--node locus. Black points mark the parameter
			combinations used for the phase portrait at $\omega=0.70$.
			The maps classify equilibrium existence and local stability,
			not basin size.
		}
		\label{fig:parameter-thresholds}
	\end{figure*}
	We next examine how operating cost, collective loss, and risk
	sensitivity modify the parameter region in which a locally
	asymptotically stable, nontrivial $D$--$S$ equilibrium exists. For
	each sampled pair $(\omega,\eta)$, where
	$\eta\in\{k,L,\gamma\}$, we computed all nontrivial roots of
	\begin{equation*}
		B(0,z;\omega,\eta)=0,
		\qquad 0<z<1,
	\end{equation*}
	and evaluated their transverse and tangential eigenvalues,
	\begin{equation*}
		\lambda_{\perp}=A(0,z;\omega,\eta),
		\qquad
		\lambda_{\parallel}
		=z(1-z)B_z(0,z;\omega,\eta).
	\end{equation*}
	A parameter pair is classified as ``present'' only when at least one
	boundary root satisfies
	$\lambda_{\perp}<0$ and $\lambda_{\parallel}<0$. The resulting regions
	therefore describe the existence of a locally asymptotically stable
	boundary equilibrium, not its basin size. The pointwise classification
	procedure is described in~\ref{app:two-parameter-folds}.
	
	Figure~\ref{fig:parameter-thresholds}(a) shows that increasing the
	operating cost $k$ shifts the lower boundary of the stable region
	toward greater protection effectiveness. The boundary is close to
	linear over the reported interval. This pattern is consistent with
	the affine dependence
	\[
	B(0,z;\omega,k)=B_0(z)-k+\omega B_1(z),
	\]
	although the boundary is not exactly linear because the equilibrium
	frequency $z$ changes along it. At the baseline value $k=0.4$, the
	lower boundary passes through
	$\omega\approx0.5254$, in agreement with SN2 in
	Fig.~\ref{fig:continuation}.
	
	Collective loss produces a more complex stability structure
	[Fig.~\ref{fig:parameter-thresholds}(b)]. Over the lower and
	intermediate part of the sampled $L$ interval, increasing $L$ extends
	the stable region toward smaller values of $\omega$. At sufficiently
	large $L$, however, the stable region is no longer separated from the
	absent region by a single threshold curve. For example, at $L=6$, the
	sampled grid detects a locally stable boundary equilibrium for
	approximately
	\[
	0.3175\leq\omega\leq0.3225
	\quad\text{and}\quad
	0.5925\leq\omega\leq1,
	\]
	with no such equilibrium detected between these intervals. Thus,
	larger collective loss can reorganize the boundary equilibrium
	structure rather than merely lower a unique effectiveness threshold.
	This pattern arises numerically because $L$ changes all
	risk-dependent terms in the boundary payoff difference, not only the
	pivotal benefit $L\omega\Psi_M$.
	
	Risk sensitivity also has a non-monotonic effect
	[Fig.~\ref{fig:parameter-thresholds}(c)]. The lower boundary of the
	stable region initially shifts toward smaller $\omega$, reaches its
	leftmost position near $\gamma\approx0.61$, and then moves toward
	greater effectiveness. This reversal is consistent with the
	saturation of
	$p(d)=1-e^{-\gamma d}$: once failure is already likely for the
	relevant group compositions, further increases in $\gamma$ produce
	smaller changes in the risk differences entering selection. The
	numerical map establishes the non-monotonic pattern over the reported
	parameter interval but does not constitute an analytical proof of its
	cause.

	\subsection{Quorum size shifts the stability threshold}
	\label{subsec:quorum-threshold}
	
	The quorum size $M$ is discrete, so we compare separate boundary
	calculations rather than treat it as a continuation parameter. For
	$N=5$ and the remaining baseline parameters, Table~\ref{tab:quorum-threshold}
	shows that the critical effectiveness increases from $0.5254$ at
	$M=2$ to $0.8321$ at $M=4$. The route to stability also changes. At
	$M=2$, the stable branch is created directly at the boundary fold
	SN2. At $M=3$ and $M=4$, a $D$--$S$ branch exists before it becomes
	stable; local asymptotic stability is acquired when the transverse
	$C$-invasion eigenvalue $\lambda_{\perp}=A(0,z)$ crosses zero.
	
	\begin{table}[t]
		\centering
		\small
		\renewcommand{\arraystretch}{1.3} 
		\caption{
			Critical protection effectiveness $\omega_c$ and the dynamical route to local asymptotic stability across discrete quorum sizes $M$.
		}
		\label{tab:quorum-threshold}
		\begin{tabular}{@{} c c c l @{}}
			\toprule
			$M$ & $\omega_c$ & $z_c$ & \textbf{Local transition} \\
			\midrule
			$2$ & $0.5254$ & $0.2718$ & Boundary saddle--node \\
			$3$ & $0.6608$ & $0.7935$ & Transverse invasion crossing \\
			$4$ & $0.8321$ & $0.9541$ & Transverse invasion crossing \\
			\bottomrule
		\end{tabular}
	\end{table}
	Changing $M$ replaces $\Psi_M$ by the risk-weighted probability of
	encountering exactly $M-1$ protective co-players. It therefore changes
	both the location and magnitude of the pivotality term in the
	simplex. For the examined values, a stricter quorum requires greater
	protection effectiveness and can change the bifurcation mechanism by
	which the boundary equilibrium becomes stable. This comparison does
	not establish monotonicity for other group sizes or parameter sets.
	
	\subsection{Numerically observed basin geometry after SN2}
	\label{subsec:basins}
	\begin{figure}[t]
		\centering
		\includegraphics[width=0.70\linewidth]
		{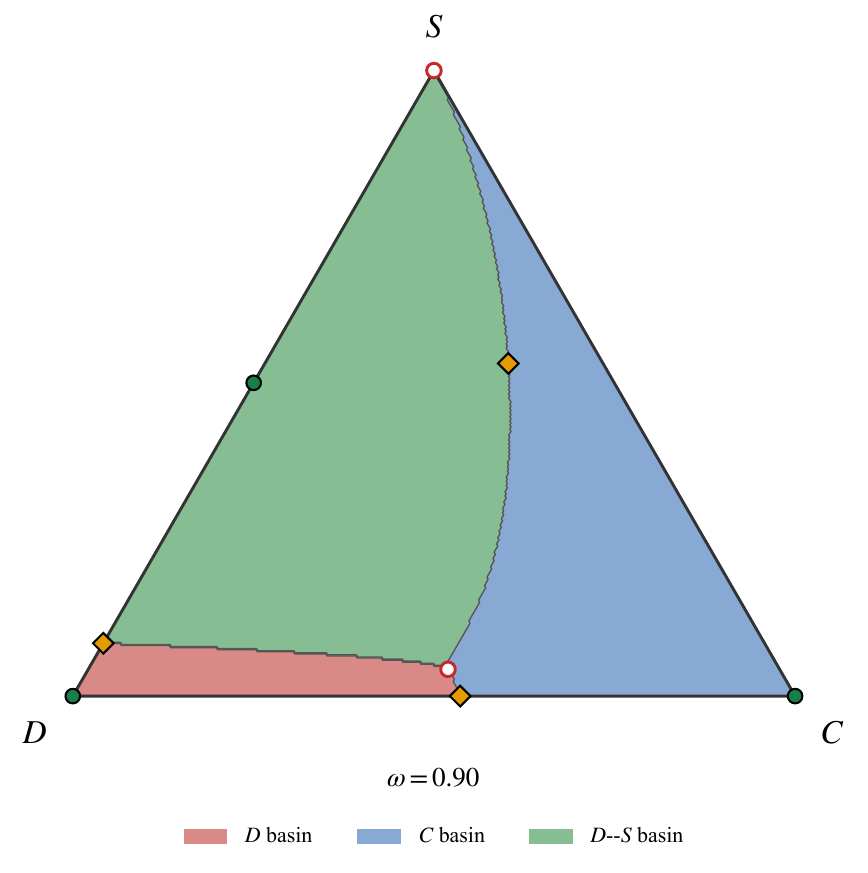}
		\caption{\textbf{Numerically observed basins at $\omega=0.90$ under the	baseline parameters.}
			Red, blue, and green initial states
			approach full defection, full cooperation, and the locally
			asymptotically stable $D$--$S$ boundary equilibrium,
			respectively. Filled green circles, orange diamonds, and open
			red circles denote locally asymptotically stable equilibria,
			saddles, and unstable equilibria. Terminal states were assigned
			only when their Euclidean distance from an independently
			computed attractor was below $10^{-6}$. The exact pure-$S$
			state remains unclassified. The colored regions are finite-grid
			numerical basin estimates rather than a global basin
			classification.
		}
		\label{fig:basin}
	\end{figure}
	Local asymptotic stability does not determine which equilibrium is
	approached from a given initial composition. We therefore integrated
	the replicator system from a uniform barycentric grid at
	$\omega=0.90$. At this parameter value, the locally asymptotically
	stable boundary equilibrium is
	\begin{equation*}
		E_{DS}(0.90)
		=
		(0,0.4992,0.5008),
	\end{equation*}
	with eigenvalues $-0.8428$ and $-0.4836$.
	
	Figure~\ref{fig:basin} shows the numerically observed basins of full
	defection, full cooperation, and $E_{DS}(0.90)$. The sampled basin
	boundaries pass through the reported saddles and are consistent with
	their stable manifolds. The protective boundary equilibrium attracts
	a positive fraction of the sampled initial conditions, but its basin
	does not contain the full-defection vertex. Under the reported
	parameters, initial compositions on the defector-heavy side of the
	observed separatrix approach full defection rather than the protective
	boundary state.
	
	The basin calculation classified $45\,450$ of the $45\,451$ initial
	states. The only unclassified state was the exact pure-$S$
	equilibrium, which remains at that unstable equilibrium and does not
	approach any of the three attractors used in the basin assignment.
	Repeating the calculation with half the integration step and,
	separately, with a longer integration horizon produced no
	classification changes among the resolved grid points. These results
	support the displayed numerical partition for the stated parameters,
	but they do not constitute a global convergence theorem.

	\section{Discussion and conclusion}
	\label{sec:discussion}
	
	This study is concerned not with whether protection can reduce
	collective risk after activation, but with how a threshold-dependent
	and non-excludable protective activity enters the evolutionary
	process. Analysis of the endogenous-risk public-goods game shows that
	threshold activation generates qualitatively different selection
	effects across population compositions. The payoff advantage of
	protective over ordinary cooperation does not increase continuously
	with the abundance of protective cooperators. Instead, it is
	concentrated in pivotal configurations where the focal individual
	completes the activation threshold. This composition dependence
	creates a dynamical separation between the rare-strategy stage and the
	established protective state. Near full defection, rare protective
	cooperators cannot activate the mechanism, so protection effectiveness
	does not alter their local invasion direction. Once protective
	cooperators have an established presence, however, the same protection
	effectiveness can substantially reshape the nonlinear selection
	gradient and create a locally asymptotically stable coexistence state
	of defectors and protective cooperators through a boundary
	saddle--node bifurcation. The appearance of this new attractor does not
	require full defection to lose its local stability, allowing the
	protective coexistence state and full defection to persist
	simultaneously. In parameter regimes where full cooperation is also
	locally asymptotically stable, all three long-run outcomes coexist on
	the strategy simplex. The resulting evolutionary picture differs from
	one based on successful invasion from rarity. A protective mechanism
	may remain stable after it has been established without being able to
	form from a small number of participants, and the state ultimately
	reached depends on the initial population composition and its
	associated basin of attraction.
	
	The dynamical results can be understood through a mechanism that links
	individual payoff differences to population-level bifurcation
	structure. Ordinary and protective cooperators make the same
	public-good contribution, so selection between them depends only on
	the operating cost continuously paid by protective cooperators and the
	expected loss avoided when they complete the activation threshold.
	Their payoff difference is therefore
	\[
	P_S-P_C=-k+L\omega\Psi_M,
	\]
	where
	\[
	\Psi_M=\mathbb{E}\!\left[p(j)\mathbb{I}_{m=M-1}\right]
	\]
	is not the unconditional probability that the group reaches the
	activation threshold. Instead, it is the risk-weighted probability
	that the focal protective cooperator becomes the $M$th participant
	and thereby activates protection. The condition $m=M-1$ ensures that
	the focal individual is indispensable for activation, while the risk
	weight $p(j)$ ensures that this pivotal role produces a benefit only
	when defectors generate a positive failure risk. This definition first
	explains why the advantage of protective cooperation depends on group
	composition. When protective cooperators are rare, the focal
	individual is unlikely to encounter $M-1$ protective co-players. As
	the population approaches the pure-protection state, the threshold is
	usually satisfied without the focal individual, while the declining
	number of defectors also drives $p(j)$ toward zero. The protective
	benefit therefore does not increase continuously with
	protective-cooperator frequency. Instead, it is concentrated in
	intermediate group compositions where failure risk and pivotality are
	simultaneously substantial. The same structure explains the barrier
	to invasion from rarity. For $M\geq2$, $\Psi_M$ scales with
	$z^{M-1}$ near full defection, and the replicator equation for the
	protective strategy contains an additional multiplicative frequency
	factor $z$. The growth effect generated by protection effectiveness
	therefore enters the dynamical system at order $z^M$ or higher and
	does not appear in the first-order Jacobian at full defection.
	Increasing $\omega$ can amplify the benefit of protection after
	activation, but it cannot change the local invasion direction of a
	rare protective cooperator near full defection. Once $z$ moves away
	from zero, pivotal configurations occur with non-negligible
	probability, allowing $L\omega\Psi_M$ to compete with the fixed cost
	$k$. On the defector--protective-cooperator boundary, this competition
	enters the zero-growth condition $B(0,z;\omega)=0$. As $\omega$
	increases, the protective benefit changes the nonlinear dependence of
	this condition on $z$. When the additional tangency condition
	$B_z(0,z;\omega)=0$ is satisfied, the boundary equilibria undergo a
	saddle--node bifurcation. The resulting stable coexistence reflects
	the feedback between risk generation and risk mitigation. More
	defectors raise the failure probability and thereby increase the
	potential value of protection. More protective cooperators suppress
	the failure probability, but their operating cost does not disappear
	as risk declines. If defectors vanish completely, then $p(0)=0$ and
	protection produces no additional benefit, allowing ordinary
	cooperators to replace protective cooperators through their cost
	advantage. This feedback explains why protective cooperation may fail
	to become established from rarity yet remain stable in coexistence
	with defection in population compositions that retain sufficient
	exposure to collective risk.
	
	Existing studies have generally approached cooperation under collective
	risk along two broad lines. Collective-risk models link public-good
	contributions directly to risk mitigation, so that reaching an aggregate
	contribution target prevents a shared loss that otherwise occurs with an
	exogenously specified probability~\citep{milinski2008collective,
		wang2009emergence,santos2011risk,chen2012risk}. Related cooperation
	models alter strategy-specific payoffs or redistribute exposure to
	collective losses through punishment~\citep{fehr2000cooperation,boyd2003evolution}, social exclusion~\citep{sasaki2013evolution}, or insurance~\citep{zhang2015insurance}. Recent research has extended these approaches
	by considering centralized monitoring and reporting, competing
	risk-sharing arrangements, adaptive governance, and coevolutionary
	feedback among cooperation, risk, and cost~\citep{he2019central,wang2025strategic,hua2026adaptive,
		xu2026adaptive,wang2026coevolutionary}. These approaches emphasize
	aggregate contributions, targeted incentives, institutional adaptation,
	or continuously changing risk environments, but they do not separate
	ordinary public-good provision from a distinct non-excludable protective
	activity activated by its own participation quorum.
	
	The present model separates failure-risk generation, public-good
	contribution, and collective protection into three related but distinct
	processes. Defection endogenously raises the failure probability,
	protective cooperators bear an additional cost beyond their public-good
	contribution, and activated protection benefits all group members,
	including defectors. Risk mitigation therefore enters evolutionary
	selection through pivotality in risk-bearing groups rather than directly
	through aggregate contributions or targeted transfers. Protective
	cooperation may consequently fail to invade because it cannot activate
	the mechanism when rare, yet form a locally asymptotically stable
	coexistence state through the creation of a new attractor at a positive
	frequency without requiring full defection to lose its local stability.
	This result extends the analysis of threshold-dependent protection beyond
	the rare-invasion criterion to local bifurcations and basin geometry. It
	also distinguishes the effectiveness of protection after activation, its
	capacity to form from a small number of participants, and its stable
	maintenance after establishment. Cross-border epidemic surveillance
	provides a relevant example because its effectiveness depends on shared
	national surveillance and reporting capacities. Improving monitoring or
	early-warning technology can increase the protective value of an
	established system, while its initial formation still depends on the
	operating cost, activation threshold, and initial participation structure.
	
	Some of the model assumptions naturally limit the broader validity of
	our conclusions. The infinite
	well-mixed population excludes spatial clustering and network
	heterogeneity. Local assortment may allow protective cooperators that
	are globally rare to satisfy the activation threshold within
	particular groups, but it may also produce spatial lock-in between
	protected and unprotected regions. The deterministic replicator
	dynamics do not capture demographic drift, mutation, or
	noise-induced transitions across basin boundaries in finite
	populations. Players are also assumed to have homogeneous contribution
	costs, operating costs, and protective capacities, while participation
	recognition and threshold activation are error free. Cost
	heterogeneity, recognition errors, and activation failure may alter
	both pivotality and local stability. In addition, the continuation
	analysis identifies local bifurcations, and the numerically observed
	basins apply only to the reported parameter ranges rather than
	constituting a global convergence result. Future research can extend
	the mechanism to complex networks, spatial lattices, and finite
	populations, examine how stochasticity, heterogeneity, and local
	assortment reshape the invasion barrier and basin structure, and allow
	the activation threshold, operating cost, and protection effectiveness
	to coevolve with population composition and environmental risk.
	
	\appendix
	
	\section{Derivation of the expected payoffs}
	\label{app:payoffs}
	
	This appendix derives the expected payoffs from the
	composition-dependent payoffs in Eq.~\eqref{eq:single-payoffs}. We
	first evaluate the risk-weighted activation terms and then assemble
	the public-good and collective-loss components.
	
	\subsection{Random matching}
	\label{app:random-matching}
	
	Among the $n=N-1$ co-players of a focal individual, the numbers
	$(i,j,m)$ of ordinary cooperators, defectors, and protective
	cooperators follow the multinomial distribution
	\begin{equation*}
		\Pr(i,j,m)
		=
		\frac{n!}{i!j!m!}
		x^iy^jz^m,
	\end{equation*}
	subject to the strict composition constraint $i+j+m=n$. Hence, for any focal strategy $\ell\in\{C,D,S\}$, the expected population-level payoff is formulated as
	\begin{equation}
		P_\ell
		=
		\sum_{\substack{i,j,m\geq0\\i+j+m=n}}
		\Pr(i,j,m)\pi_\ell(i,j,m).
		\label{eq:general-expected-payoff}
	\end{equation}
	All expressions below are exact analytical evaluations of this finite
	multinomial sum.
	
	\subsection{Risk-weighted activation terms}
	\label{app:quorum-terms}
	
	Starting from Eq.~\eqref{eq:quorum-risk-expectations}, the
	risk-weighted activation term for a focal contributor unfolds as
	\[
	T_h
	=
	\sum_{m=h}^{n}
	\sum_{j=0}^{n-m}
	\frac{n!}{(n-m-j)!j!m!}
	x^{n-m-j}y^jz^m p(j).
	\]
	Conversely, for a focal defector, its own strategy intrinsically changes the total defector count from $j$ to $j+1$, yielding
	\[
	T_h^D
	=
	\sum_{m=h}^{n}
	\sum_{j=0}^{n-m}
	\frac{n!}{(n-m-j)!j!m!}
	x^{n-m-j}y^jz^m p(j+1).
	\]
	
	Fixing $m$ leaves $n-m$ positions occupied by ordinary cooperators and defectors. Summing the corresponding multinomial weights over these remaining positions gives
	\begin{equation}
		\sum_{j=0}^{n-m}
		\frac{n!}{(n-m-j)!j!m!}
		x^{n-m-j}y^jz^m
		=
		\binom{n}{m}z^m(1-z)^{n-m}.
		\label{eq:app-unweighted-sum}
	\end{equation}
	If we include the failure-survival factor $\rho^j$ in the summation instead, we obtain
	\begin{equation}
		\sum_{j=0}^{n-m}
		\frac{n!}{(n-m-j)!j!m!}
		x^{n-m-j}y^jz^m\rho^j
		=
		\binom{n}{m}z^m(x+\rho y)^{n-m}.
		\label{eq:app-weighted-sum}
	\end{equation}
	Because the failure probability is defined as $p(j)=1-\rho^j$, subtracting
	Eq.~\eqref{eq:app-weighted-sum} from
	Eq.~\eqref{eq:app-unweighted-sum} and summing over all states satisfying the threshold $m\geq h$ explicitly yields
	\begin{equation}
		T_h(x,y,z)
		=
		\sum_{m=h}^{n}
		\binom{n}{m}z^m
		\left[
		(1-z)^{n-m}
		-(x+\rho y)^{n-m}
		\right].
		\label{eq:Th}
	\end{equation}
	
	Similarly, for a focal defector, the incremented risk follows $p(j+1)=1-\rho^{j+1}=1-\rho\rho^j$. Applying the identical algebraic calculation gives
	\begin{equation}
		T_h^D(x,y,z)
		=
		\sum_{m=h}^{n}
		\binom{n}{m}z^m
		\left[
		(1-z)^{n-m}
		-\rho(x+\rho y)^{n-m}
		\right].
		\label{eq:ThD}
	\end{equation}
	Equations~\eqref{eq:Th} and \eqref{eq:ThD} serve as the finite-sum
	representations of the expectations previously defined in Eq.~\eqref{eq:quorum-risk-expectations}.
	
	\subsection{Unprotected failure probabilities}
	\label{app:unprotected-risks}
	
	Setting $h=0$ effectively removes the activation filter. By applying the binomial
	theorem to Eq.~\eqref{eq:Th}, the baseline unprotected risk simplifies to
	\[
	\begin{aligned}
		T_0
		&=
		\sum_{m=0}^{n}
		\binom{n}{m}z^m(1-z)^{n-m}
		-
		\sum_{m=0}^{n}
		\binom{n}{m}z^m(x+\rho y)^{n-m}
		\\
		&=
		1-(z+x+\rho y)^n
		=
		1-q^n.
	\end{aligned}
	\]
	The corresponding expression for a focal defector is straightforwardly evaluated as $T_0^D = 1-\rho q^n$. Equivalently, by substituting $j\sim\operatorname{Binomial}(n,y)$, we can confirm that the expectation evaluates to $\mathbb E[\rho^j] = (1-y+\rho y)^n = q^n$. It naturally follows that the additional expected failure probability generated by the focal defector is exactly $T_0^D-T_0 = (1-\rho)q^n > 0$.
	
	\subsection{Assembly of the expected payoffs}
	\label{app:payoff-assembly}
	
	In a realized group, the expected number of contributing co-players is $\mathbb E[i+m]=n(x+z)$. Because a focal $C$- or $S$-player also contributes to the public good, their inclusion raises this expectation to $\mathbb E[i+m+1]=1+n(x+z)$.
	
	For a focal contributor subject to a generic co-player threshold $h$, the expected residual failure probability decomposes as
	\begin{equation}
		\begin{aligned}
			\mathbb E\!\left[
			p(j)
			\left(
			1-\omega\mathbb I_{m\geq h}
			\right)
			\right]
			&=
			\mathbb E[p(j)]
			-
			\omega
			\mathbb E\!\left[
			p(j)\mathbb I_{m\geq h}
			\right]
			\\
			&=
			T_0-\omega T_h.
		\end{aligned}
		\label{eq:app-residual-risk}
	\end{equation}
	For a focal defector, the corresponding residual risk expression is analogous, yielding $T_0^D-\omega T_h^D$.
	
	A focal ordinary cooperator explicitly requires $m\geq M$ for activation. Combining the expected public-good return with the residual risk derived in Eq.~\eqref{eq:app-residual-risk} gives
	\[
	P_C
	=
	\frac{rc[1+n(x+z)]}{N}-c
	-L[T_0-\omega T_M].
	\]
	Conversely, a focal defector faces the same activation condition but strictly alters both the contribution count and the failure probability:
	\[
	P_D
	=
	\frac{rcn(x+z)}{N}
	-L[T_0^D-\omega T_M^D].
	\]
	Finally, a focal protective cooperator intrinsically counts toward the required threshold, therefore needing only $M-1$ protective co-players to trigger the mechanism:
	\[
	P_S
	=
	\frac{rc[1+n(x+z)]}{N}-c-k
	-L[T_0-\omega T_{M-1}].
	\]
	These derivations successfully reproduce the expected payoffs in Eq.~\eqref{eq:PS} presented in the main text.
	
	\section{Pivotality identity and closed payoff differences}
	\label{app:payoff-differences}
	
	\subsection{Risk-weighted pivotality}
	\label{app:pivotality}
	
	The terms $T_{M-1}$ and $T_M$ differ only in the specific configuration where $m=M-1$. Therefore, their difference directly isolates the pivotal probability:
	\begin{equation}
		\begin{aligned}
			T_{M-1}-T_M
			&=
			\mathbb E\!\left[
			p(j)\mathbb I_{m=M-1}
			\right]
			\\
			&=
			\binom{n}{M-1}z^{M-1}
			\left[
			(1-z)^{n-M+1}
			-(x+\rho y)^{n-M+1}
			\right]
			\\
			&=
			\Psi_M.
		\end{aligned}
	\end{equation}
	This formally proves the expansion of $\Psi_M$ in Eq.~\eqref{eq:pivotal-expectation}. Furthermore, subtracting the expected payoff of ordinary cooperation from that of protective cooperation yields
	\begin{equation}
		P_S-P_C
		=
		-k+L\omega(T_{M-1}-T_M)
		=
		-k+L\omega\Psi_M,
		\label{eq:app-pivotal-identity}
	\end{equation}
	which structurally proves Eq.~\eqref{eq:pivotal-identity}.
	
	\subsection{Closed payoff differences}
	\label{app:closed-differences}
	
	The survival-weighted activation term $R_M$ initially defined in Section~\ref{subsec:selection-gradients} admits the finite-sum representation
	\[
	R_M(x,y,z)
	=
	\sum_{m=M}^{n}
	\binom{n}{m}
	z^m(x+\rho y)^{n-m}.
	\]
	Indeed, fixing $m$ and summing the survival factor $\rho^j$ over the remaining positions occupied by ordinary cooperators and defectors directly gives $(x+\rho y)^{n-m}$. By subtracting Eq.~\eqref{eq:Th} from Eq.~\eqref{eq:ThD} evaluated at $h=M$, we obtain the risk difference
	\[
	\begin{aligned}
		T_M^D-T_M
		&=
		(1-\rho)
		\sum_{m=M}^{n}
		\binom{n}{m}
		z^m(x+\rho y)^{n-m}
		\\
		&=
		(1-\rho)R_M.
	\end{aligned}
	\]
	For the gradient $A=P_C-P_D$, the public-good terms naturally reduce to the baseline margin $a=rc/N-c$, while the collective-loss terms expand to
	\[
	\begin{aligned}
		A
		&=
		a-L
		\left[
		(T_0-\omega T_M)
		-
		(T_0^D-\omega T_M^D)
		\right]
		\\
		&=
		a+L(1-\rho)
		\left[
		q^n-\omega R_M
		\right].
	\end{aligned}
	\]
	After explicitly substituting the simplex constraint $y=1-x-z$, this derivation yields Eq.~\eqref{eq:A}. Finally, the gradient $B$ is recovered via
	\[
	\begin{aligned}
		B
		&=
		P_S-P_D
		=
		(P_C-P_D)+(P_S-P_C)
		\\
		&=
		A-k+L\omega\Psi_M,
	\end{aligned}
	\]
	which proves Eq.~\eqref{eq:B}.
	
	\section{Boundary algebra and Jacobian calculations}
	\label{app:boundary-algebra}
	
	\subsection{General Jacobian}
	\label{app:jacobian}
	
	For notational readability, the explicit dependence of the functions $A$, $B$, $F$, and $G$
	on the efficacy parameter $\omega$ is dynamically suppressed in this section. We rewrite the planar system in Eq.~\eqref{eq:planar} as the vector field
	\[
	\begin{aligned}
		F(x,z)
		&=
		x\left[
		(1-x)A(x,z)-zB(x,z)
		\right],
		\\
		G(x,z)
		&=
		z\left[
		(1-z)B(x,z)-xA(x,z)
		\right].
	\end{aligned}
	\]
	Direct partial differentiation of these components yields
	\begin{equation}
		\begin{aligned}
			F_x
			&=
			(1-2x)A+x(1-x)A_x-zB-xzB_x,
			\\
			F_z
			&=
			x\left[
			(1-x)A_z-B-zB_z
			\right],
			\\
			G_x
			&=
			z\left[
			(1-z)B_x-A-xA_x
			\right],
			\\
			G_z
			&=
			(1-2z)B-xA+z(1-z)B_z-xzA_z.
		\end{aligned}
		\label{eq:general-jacobian-components}
	\end{equation}
	Thus, the continuous-time Jacobian matrix is defined as
	\begin{equation}
		J(x,z)
		=
		\begin{pmatrix}
			F_x&F_z\\
			G_x&G_z
		\end{pmatrix}.
		\label{eq:general-jacobian}
	\end{equation}
	
	\subsection{Jacobians at the homogeneous equilibria}
	\label{app:homogeneous-jacobians}
	
	At the full-defection state, the reduced coordinates reside at $(x,z)=(0,0)$, yielding the base selection gradients $A(0,0)=A_D$ and $B(0,0)=A_D-k$. Substitution into the general components constructs the diagonal Jacobian matrix
	\begin{equation}
		J(0,0)
		=
		\begin{pmatrix}
			A_D&0\\
			0&A_D-k
		\end{pmatrix},
		\label{eq:D-jacobian}
	\end{equation}
	whose diagonal entries inherently provide the invasion eigenvalues reported in Eq.~\eqref{eq:D-eigen}.
	
	Similarly, at the full-cooperation state, the coordinates shift to $(x,z)=(1,0)$ with $A(1,0)=A_C$ and $B(1,0)=A_C-k$. Hence, the corresponding Jacobian evaluates to
	\begin{equation}
		J(1,0)
		=
		\begin{pmatrix}
			-A_C&k-A_C\\
			0&-k
		\end{pmatrix},
		\label{eq:C-jacobian}
	\end{equation}
	whose structural diagonal entries reproduce Eq.~\eqref{eq:C-eigenvalues}.
	
	Finally, at full protective cooperation, $(x,z)=(0,1)$. Defining $A_S = a+L(1-\rho)(1-\omega)$, and observing that $\Psi_M=0$ in the total absence of defectors, yields $B(0,1)=A_S-k$. The resultant Jacobian becomes
	\begin{equation}
		J(0,1)
		=
		\begin{pmatrix}
			k&0\\
			-A_S&k-A_S
		\end{pmatrix}.
		\label{eq:S-jacobian}
	\end{equation}
	Its exact eigenvalues are determined to be $k$ and
	$k-a-L(1-\rho)(1-\omega)$, strictly aligning with Eq.~\eqref{eq:S-eigenvalues}.
	
	\subsection{Nonhyperbolic homogeneous states}
	\label{app:nonhyperbolic-vertices}
	
	The zero-eigenvalue cases at the homogeneous states fundamentally require a boundary-flow argument to rigorously ascertain their stability. 
	
	On the invariant $C$--$D$ edge, the restricted gradient defined as
	\[
	A_{CD}(x)
	=
	a+L(1-\rho)
	\left[
	\rho+(1-\rho)x
	\right]^n
	\]
	is a strictly increasing function of $x$. If $A_D=A_{CD}(0)=0$, then $A_{CD}(x)>0$ for every sufficiently small perturbation $x>0$. Since the local dynamic flow satisfies $\dot x=x(1-x)A_{CD}(x)>0$, nearby states strictly move away from full defection. Thus, the full-defection state is classified as nonhyperbolic and structurally unstable when $A_D=0$. Conversely, when $A_D=k$, the complementary eigenvalue is strictly positive ($\lambda_C^D=k>0$), rendering the state unstable regardless of the zero eigenvalue.
	
	If $A_C=A_{CD}(1)=0$, the strict monotonicity implies $A_{CD}(x)<0$ for any $x<1$ sufficiently close to unity. Consequently, $\dot x=x(1-x)A_{CD}(x)<0$, forcing nearby edge states to move away from full cooperation. The state is therefore topologically nonhyperbolic and unstable when $A_C=0$.
	
	At full protective cooperation, the invasion eigenvalue of an ordinary cooperator is $\lambda_C^S=k>0$ across every maintained parameter regime. Hence, this state fundamentally remains unstable even when its second eigenvalue vanishes.
	
	\subsection{The $C$--$D$ edge}
	\label{app:boundary-stability}
	
	On the boundary $z=0$, we have $y=1-x$, and the activation terms structurally vanish because $M\geq2$. Equation~\eqref{eq:A} therefore reduces to
	\[
	A_{CD}(x)
	:=
	A(x,0)
	=
	a+L(1-\rho)
	\left[
	\rho+(1-\rho)x
	\right]^n.
	\]
	Its derivative with respect to $x$ is strictly positive, explicitly evaluating to
	\[
	A_{CD}'(x)
	=
	nL(1-\rho)^2
	\left[
	\rho+(1-\rho)x
	\right]^{n-1}
	>
	0.
	\]
	Because the endpoints naturally satisfy $A_{CD}(0)=A_D$ and $A_{CD}(1)=A_C$, a unique nontrivial edge equilibrium emerges precisely when $A_D<0<A_C$. Analytically solving the condition $A_{CD}(x)=0$ yields the coordinate
	\[
	x_{CD}
	=
	\frac{
		\left[
		-a/\bigl(L(1-\rho)\bigr)
		\right]^{1/n}
		-\rho
	}{
		1-\rho
	}.
	\]
	At this precisely located equilibrium, the gradient conditions $A(x_{CD},0)=0$ and $B(x_{CD},0)=-k$ hold. The resulting Jacobian takes an upper triangular form:
	\begin{equation}
		J(x_{CD},0)
		=
		\begin{pmatrix}
			x_{CD}(1-x_{CD})A_x(x_{CD},0)&\ast\\
			0&-k
		\end{pmatrix}.
		\label{eq:CD-jacobian}
	\end{equation}
	The tangential eigenvalue is strictly positive while the transverse eigenvalue is negative, verifying that the interior edge equilibrium $E_{CD}$ constitutes a topological saddle.
	
	\subsection{The excluded zero-cost limit}
	\label{app:zero-cost}
	
	The standard formulation of the model assumes a strictly positive operating cost ($k>0$). In the hypothetical limit where $k=0$, the system constrained to the $C$--$S$ edge features $y=0$, $p(0)=0$, and $\Psi_M=0$. Therefore, the payoff gap closes ($P_S-P_C=0$), implying $B=A$. Given the simplex identity $x+z=1$ along this boundary, the planar vector field completely vanishes:
	\[
	\begin{aligned}
		F(x,z)
		&=
		x(1-x-z)A
		=
		0,
		\\
		G(x,z)
		&=
		z(1-x-z)A
		=
		0.
	\end{aligned}
	\]
	Consequently, every singular point situated on the $C$--$S$ edge acts as an equilibrium when $k=0$. This structurally creates a nonhyperbolic continuum, although its localized transverse stability concerning defector invasions may selectively vary along the edge itself.
	
	\subsection{The $D$--$S$ edge}
	\label{app:DS-jacobian}
	
	At any nontrivial equilibrium satisfying the root condition $B(0,z_{DS};\omega)=0$, the general Jacobian matrix derived in Eq.~\eqref{eq:general-jacobian-components} simplifies to
	\begin{equation}
		J(0,z_{DS};\omega)
		=
		\begin{pmatrix}
			A(0,z_{DS};\omega)&0\\
			\ast&
			z_{DS}(1-z_{DS})B_z(0,z_{DS};\omega)
		\end{pmatrix}.
		\label{eq:DS-jacobian}
	\end{equation}
	Its diagonal entries explicitly supply the transverse and tangential eigenvalues previously presented in Eq.~\eqref{eq:DS-eigenvalues}.
	
	\subsection{Boundary saddle--node reduction}
	\label{app:boundary-fold-reduction}
	
	Holding all structural parameters other than $\omega$ completely fixed, the dynamics rigorously restricted to the $D$--$S$ edge are governed by the scalar field
	\begin{equation}
		\dot z
		=
		\mathcal G(z,\omega)
		:=
		z(1-z)B(0,z;\omega).
		\label{eq:DS-restricted-field}
	\end{equation}
	At a nontrivial boundary equilibrium where $B(0,z;\omega)=0$, the first derivative inherently simplifies to $\mathcal G_z = z(1-z)B_z(0,z;\omega)$. The fundamental fold condition $\mathcal G_z=0$ is therefore algebraically equivalent to $B_z=0$.
	
	Evaluating a candidate fold that simultaneously satisfies $B=B_z=0$, the parametric and higher-order spatial derivatives expand as
	\[
	\begin{aligned}
		\mathcal G_\omega
		&=
		z(1-z)B_\omega(0,z;\omega),
		\\
		\mathcal G_{zz}
		&=
		z(1-z)B_{zz}(0,z;\omega).
	\end{aligned}
	\]
	Because the interior term $z(1-z)$ is strictly positive, the scalar nondegeneracy transversality conditions $\mathcal G_\omega\neq0$ and $\mathcal G_{zz}\neq0$ are precisely equivalent to $B_\omega(0,z;\omega)\neq0$ and $B_{zz}(0,z;\omega)\neq0$.
	
	Furthermore, the secondary eigenvalue embedded in the full planar system is $\lambda_\perp=A(0,z;\omega)$. The non-zero requirement $\lambda_\perp\neq0$ guarantees that the tangential zero eigenvalue remains geometrically simple. If it specifically evaluates to $\lambda_\perp<0$, the directional manifold transverse to the invariant edge behaves as locally attracting.
	
	\subsection{Interior equilibria}
	\label{app:interior-jacobian}
	
	At a fully polymorphic interior equilibrium, the selection gradients vanish ($A=B=0$). Substituting these conditions directly into Eq.~\eqref{eq:general-jacobian-components} gives the specific Jacobian
	\begin{equation}
		J_*
		=
		\begin{pmatrix}
			x^*[(1-x^*)A_x-z^*B_x]
			&
			x^*[(1-x^*)A_z-z^*B_z]
			\\
			z^*[(1-z^*)B_x-x^*A_x]
			&
			z^*[(1-z^*)B_z-x^*A_z]
		\end{pmatrix}_{E_*}.
		\label{eq:app-interior-J}
	\end{equation}
	Remarkably, this dense matrix gracefully factorizes into the product structure
	\begin{equation}
		J_*
		=
		\begin{pmatrix}
			x^*&0\\
			0&z^*
		\end{pmatrix}
		\begin{pmatrix}
			1-x^*&-z^*\\
			-x^*&1-z^*
		\end{pmatrix}
		\begin{pmatrix}
			A_x&A_z\\
			B_x&B_z
		\end{pmatrix}_{E_*}.
		\label{eq:Jfactor}
	\end{equation}
	Calculating the determinant of the central matrix yields $(1-x^*)(1-z^*)-x^*z^* = 1-x^*-z^* = y^*$. It directly follows that the determinant of the entire system simplifies to
	\begin{equation}
		\det J_*
		=
		x^*y^*z^*
		\left(
		A_xB_z-A_zB_x
		\right)_{E_*},
		\label{eq:app-Jdet}
	\end{equation}
	which structurally proves Eq.~\eqref{eq:Jdet}. Concurrently, its scalar trace is derived as
	\begin{equation}
		\begin{aligned}
			\operatorname{tr}J_*
			={}&
			x^*
			\left[
			(1-x^*)A_x-z^*B_x
			\right]_{E_*}
			\\
			&+
			z^*
			\left[
			(1-z^*)B_z-x^*A_z
			\right]_{E_*}.
		\end{aligned}
		\label{eq:Jtrace}
	\end{equation}
	For a hyperbolic planar equilibrium, applying the Routh--Hurwitz criteria ensures local asymptotic stability if and only if $\det J_*>0$ and $\operatorname{tr}J_*<0$. A negative determinant ($\det J_*<0$) mathematically enforces two real eigenvalues possessing opposite signs, thus defining a saddle point.
	
	\section{Numerical branch tracking and bifurcation diagnostics}
	\label{app:numerical-continuation}
	
	The computational analysis utilized Python, employing NumPy and SciPy for high-precision root finding and linear algebra operations, alongside pandas for tabular data management and Matplotlib for robust visual rendering. Crucially, all equilibrium calculations were formulated strictly from the closed payoff differences previously defined in Eqs.~\eqref{eq:A} and \eqref{eq:B}. Therefore, the procedure entirely circumvented the need for any Monte Carlo approximation of the randomly matched group compositions.
	
	\subsection{Parameter-stepped equilibrium searches}
	\label{app:equilibrium-searches}
	
	For each parametrically fixed value of the protection effectiveness $\omega$, the coordinates of the interior equilibria were systematically extracted by solving the nonlinear system
	\[
	\begin{aligned}
		A(x,z;\omega)&=0,\\
		B(x,z;\omega)&=0,
	\end{aligned}
	\]
	subject to the open simplex bounds $x>0, z>0,$ and $x+z<1$. This continuous system was computationally resolved utilizing the robust Powell hybrid method with a strict solver tolerance of $10^{-11}$. The initial numerical guesses were meticulously sampled from the admissible points of a $13\times13$ grid, whose one-dimensional coordinates independently ranged from $0.025$ to $0.925$. A candidate solution was definitively retained only when all three evaluated strategy frequencies exceeded a margin of $10^{-8}$ and the final residual norm successfully fell below $10^{-8}$. Furthermore, any solutions geometrically separated by less than $10^{-6}$ in Euclidean distance were algorithmically classified as duplicate roots.
	
	Nontrivial equilibria securely located on the invariant $D$--$S$ edge rigorously satisfy the boundary condition
	\begin{equation}
		B(0,z;\omega)=0,
		\qquad
		0<z<1.
		\label{eq:continuation-DS}
	\end{equation}
	To construct the precise one-parameter branch diagram, the entire valid interval $(10^{-7},1-10^{-7})$ was initially sampled at $1800$ uniformly spaced points. Every identified sign-changing bracket was subsequently refined applying Brent's robust scalar method, with clustered roots separated by less than $10^{-5}$ being appropriately merged. Because standard sign-change routines fail to reliably detect a double root located exactly at a fold transition, the precise saddle--node coordinates were mathematically determined separate from this scan by solving the augmented equations described below.
	
	The visualization in Figure~\ref{fig:continuation} synthesized data spanning $\omega=0, 0.002, \ldots, 1$, explicitly supplemented by the independently refined bifurcation values. At each discrete value of $\omega$, the algorithm executed both the comprehensive interior seed grid and the boundary scan. The methodology is therefore most accurately characterized as a highly granular parameter-stepped root search coupled with numerical branch matching, distinguishing it from pseudo-arclength continuation algorithms. Accordingly, the reported topological branch set remains conditional on the explicitly stated search grids and strict tolerances, thereby refraining from asserting an exhaustive global mathematical enumeration of all hypothetically possible equilibria.
	
	The reduced planar Jacobian matrix was systematically evaluated via centered finite differences using a highly resolved step size of $2\cdot10^{-6}$. An identified equilibrium was definitively classified as locally asymptotically stable when both complex eigenvalue real parts fell below the strict threshold $-10^{-7}$, and as fully unstable when both mutually exceeded $10^{-7}$. It was geometrically classified as a saddle point whenever the product of the two eigenvalue real parts fell beneath $-10^{-10}$. Any residual anomalous cases manifesting a near-zero eigenvalue were appropriately flagged as nonhyperbolic and rigorously examined utilizing the augmented bifurcation transversality conditions.
	
	\subsection{Interior saddle--node diagnostics}
	\label{app:interior-fold-detection}
	
	At a verified interior equilibrium, the determinant factorization previously detailed in Eq.~\eqref{eq:Jdet} provides the compact representation $\det J = xyz\det D(A,B)$, where the structural submatrix is defined as
	\[
	D(A,B)
	=
	\begin{pmatrix}
		A_x & A_z\\
		B_x & B_z
	\end{pmatrix}.
	\]
	Because the state variables strictly satisfy $x,y,z>0$ within the simplex interior, the singular condition $\det J=0$ mathematically reduces to the exact equivalence $\det D(A,B)=0$. Consequently, the precise location of the interior saddle--node SN1 was isolated by solving the augmented algebraic system
	\begin{equation}
		\begin{aligned}
			A(x,z;\omega)&=0,\\
			B(x,z;\omega)&=0,\\
			\det D(A,B)&=0.
		\end{aligned}
		\label{eq:interior-fold-system}
	\end{equation}
	Centered finite differences executed with a step size of $2\cdot10^{-5}$ were deployed to highly accurately evaluate the partial derivatives entering $D(A,B)$.
	
	Solving this augmented system definitively identifies the critical transition point at
	\[
	(x^*,y^*,z^*,\omega_{\mathrm{SN1}})
	=
	(0.2994,0.4272,0.2734,0.2882).
	\]
	Evaluated exactly at this coordinate, the planar Jacobian structural matrix exhibits one simple zero eigenvalue alongside a singular non-zero positive eigenvalue $\lambda_{\mathrm{nz}}=0.5299>0$.
	
	To perform the necessary normal form reductions, let $v_0$ and $w_0$ denote the right and left null vectors of the evaluated Jacobian at SN1, strictly normalized to satisfy $w_0^{\mathsf T}v_0=1$. For the generalized planar vector field $f(X,\omega)$, defining the state vector $X=(x,z)^{\mathsf T}$, the canonical saddle--node nondegeneracy coefficients were analytically evaluated as
	\[
	\alpha
	=
	w_0^{\mathsf T}f_\omega,
	\qquad
	\beta
	=
	\frac{1}{2}
	w_0^{\mathsf T}D_X^2f[v_0,v_0].
	\]
	High-precision centered finite differences utilizing a discrete step of $10^{-4}$ computed the parameters as
	\begin{equation}
		\alpha_{\mathrm{SN1}}=-0.2871,
		\qquad
		\beta_{\mathrm{SN1}}=0.4826.
		\label{eq:SN1-coefficients}
	\end{equation}
	Both critical coefficients are strictly non-zero, unequivocally confirming that SN1 constitutes a structurally nondegenerate saddle--node bifurcation. Moreover, calculating their ratio yields $-\alpha_{\mathrm{SN1}}/\beta_{\mathrm{SN1}}>0$, dictating that the two distinct local equilibrium branches physically exist exclusively for parameter ranges $\omega>\omega_{\mathrm{SN1}}$. The previously evaluated positive non-zero eigenvalue geometrically classifies the emerging equilibrium pair as a saddle and a highly unstable node.
	
	To ensure computational rigor, the coefficient evaluation was iteratively repeated with successively refined finite-difference steps of $4\cdot10^{-4}$, $2\cdot10^{-4}$, $5\cdot10^{-5}$, and $2.5\cdot10^{-5}$. All subsequent refinements robustly preserved the exact four-decimal values originally reported in Eq.~\eqref{eq:SN1-coefficients}.
	
	\subsection{Boundary saddle--node and edge-transcritical diagnostics}
	\label{app:fold-detection}
	
	As established, the nontrivial boundary equilibria satisfy Eq.~\eqref{eq:continuation-DS}. The critical boundary saddle--node transition SN2 was highly refined by computationally solving the tangent system
	\begin{equation}
		B(0,z;\omega)=0,
		\qquad
		B_z(0,z;\omega)=0.
		\label{eq:boundary-fold-system}
	\end{equation}
	The partial derivative $B_z$ was reliably approximated via centered finite differences with a step of $2\cdot10^{-5}$. The solver outputs the resulting exact fold geometry at
	\[
	(z^*,\omega_{\mathrm{SN2}})
	=
	(0.2718,0.5254).
	\]
	Operating at this exact topological point, the full suite of boundary transversality diagnostics evaluates to
	\[
	\begin{aligned}
		B(0,z^*;\omega_{\mathrm{SN2}})&=0,\\
		B_z(0,z^*;\omega_{\mathrm{SN2}})&=0,\\
		A(0,z^*;\omega_{\mathrm{SN2}})&=-0.469,\\
		B_\omega(0,z^*;\omega_{\mathrm{SN2}})&=1.5511,\\
		B_{zz}(0,z^*;\omega_{\mathrm{SN2}})&=-14.6775.
	\end{aligned}
	\]
	The definitively non-zero value of $A$ confirms that the transverse eigenvalue does not simultaneously vanish at SN2, while the strictly non-zero magnitudes of $B_\omega$ and $B_{zz}$ unequivocally establish the nondegeneracy of the restricted boundary fold.
	
	For the corresponding scalar edge vector field $\mathcal G(z,\omega) = z(1-z)B(0,z;\omega)$, evaluating the centered differences with a step of $10^{-4}$ provides the fold coefficients
	\[
	\alpha_{\mathrm{SN2}}
	=
	\mathcal G_\omega
	=
	0.307,
	\qquad
	\beta_{\mathrm{SN2}}
	=
	\frac{1}{2}\mathcal G_{zz}
	=
	-1.4525.
	\]
	Because the ratio dictates $-\alpha_{\mathrm{SN2}}/\beta_{\mathrm{SN2}}>0$, the two interconnected boundary branches similarly exist purely for $\omega>\omega_{\mathrm{SN2}}$. Their respective local topological classifications were decisively determined by evaluating the eigenvalues $\lambda_{\perp}=A(0,z;\omega)$ and $\lambda_{\parallel} = z(1-z)B_z(0,z;\omega)$. The lower-$z$ topological branch manifests with $\lambda_{\parallel}>0$ and $\lambda_{\perp}<0$, inherently rendering it a saddle point. Conversely, the upper-$z$ topological branch possesses strictly negative evaluations for both eigenvalues, guaranteeing it remains locally asymptotically stable throughout the full simplex.
	
	Subsequently, the nontrivial unstable boundary branch geometrically intersects the pure-$S$ boundary branch at the critical threshold $\omega_{\mathrm{TC}} = 0.6880820$. At this structural collision point, the transversality diagnostics yield
	\[
	\begin{aligned}
		B(0,1;\omega_{\mathrm{TC}}) &= 0,\\
		B_z(0,1;\omega_{\mathrm{TC}}) &= 2.8328,\\ 
		B_\omega(0,1;\omega_{\mathrm{TC}}) &= -3.0136.
	\end{aligned}
	\]
	The pure-$S$ equilibrium trivially exists for every maintained value of $\omega$, and the nontrivial boundary branch mathematically crosses it in a purely transverse manner. This collision is therefore rigorously identified as an edge transcritical bifurcation. Furthermore, because its secondary planar eigenvalue evaluates to $0.4000>0$, the bifurcation fundamentally does not create a locally attracting equilibrium within the broader bounds of the full simplex.
	
	\subsection{Discrete activation-threshold comparison}
	\label{app:quorum-comparison}
	
	Because the activation threshold $M$ is inherently a discrete parameter, it was methodically examined through separate specialized boundary calculations. For the case $M=2$, the locally asymptotically stable boundary branch is structurally created exactly at the saddle--node previously defined by Eq.~\eqref{eq:boundary-fold-system}. For higher configurations $M=3$ and $M=4$, a nontrivial boundary branch mathematically exists before it structurally becomes locally asymptotically stable. The formally reported stability thresholds governing these transitions were found by solving the coupled root system
	\[
	\begin{aligned}
		B(0,z;\omega)&=0,\\
		A(0,z;\omega)&=0.
	\end{aligned}
	\]
	These paired equations successfully locate a zero transverse ordinary-cooperator invasion eigenvalue along an existing contiguous $D$--$S$ branch.
	
	For the configuration $M=3$, the computational solver identifies the threshold at $(z_c,\omega_c) = (0.7935, 0.6608)$, while the corresponding tangential eigenvalue evaluates robustly to $-0.1904$. Similarly, for $M=4$, the system locates the threshold at $(z_c,\omega_c) = (0.9541, 0.8321)$, with a stabilizing tangential eigenvalue of $-0.3502$. Thus, only the specific transverse eigenvalue vanishes at these two evaluated points. Geometrically, they are most accurately classified as transverse stability crossings rather than canonical boundary saddle--node bifurcations. Assigning a more rigorously specific codimension-one bifurcation type to these topological transitions would inherently require an additional specialized normal-form calculation paired with the explicit structural tracking of the associated interior branch.
	
	\subsection{Pointwise classification of two-parameter maps}
	\label{app:two-parameter-folds}
	
	The comprehensive two-parameter phase maps presented in Fig.~\ref{fig:parameter-thresholds} were systematically constructed by independently classifying every sampled parameter pair across the predefined grids. The protection effectiveness was finely sampled at $401$ equally spaced intervals spanning $0\leq\omega\leq1$. Concurrently, each targeted secondary parameter was strictly sampled at $241$ uniformly spaced values traversing the respective ranges $0.15\leq k\leq0.75$, $2.5\leq L\leq6$, and $0.35\leq\gamma\leq2.5$.
	
	Restricting the analysis to the $D$--$S$ edge, the closed payoff differences organically manifest as polynomials in $z$ while remaining strictly affine in $\omega$. For each parametrically fixed value of the selected secondary parameter $\eta\in\{k,L,\gamma\}$, these differences were algebraically restructured as
	\[
	\begin{aligned}
		A(0,z;\omega,\eta)
		&=
		A_0(z;\eta)+\omega A_1(z;\eta),\\
		B(0,z;\omega,\eta)
		&=
		B_0(z;\eta)+\omega B_1(z;\eta).
	\end{aligned}
	\]
	All complex and real roots of the resulting unified degree-$(N-1)$ polynomial $B(0,z;\omega,\eta)$ were exhaustively computed at every singularly sampled parameter pair. An analytically candidate root was strictly retained exclusively when its imaginary component fell beneath the solver threshold $10^{-8}$, its spatial bounds satisfied $10^{-9}<z<1-10^{-9}$, and its optimally scaled polynomial residual remained firmly below $10^{-7}$.
	
	For each definitively retained boundary root, its local stability characterization within the complete full simplex was robustly evaluated via the eigenvalues $\lambda_{\perp} = A(0,z;\omega,\eta)$ and $\lambda_{\parallel} = z(1-z)B_z(0,z;\omega,\eta)$. A discretely sampled parameter pair was algorithmically classified as ``present'' whenever at least one nontrivial boundary root simultaneously satisfied the rigorous stability conditions $\lambda_{\perp}<-10^{-7}$ and $\lambda_{\parallel}<-10^{-7}$. Otherwise, it was broadly classified as ``absent''. This highly granular pointwise diagnostic procedure intrinsically allows for the geometric existence of multiple stable intervals at an identical secondary-parameter value, thereby bypassing any restrictive assumptions that the stable topological region rigidly occupies the entire half-plane situated exclusively on one side of a selected fold.
	
	The distinct solid contours graphically overlaid in Fig.~\ref{fig:parameter-thresholds} were computationally extracted directly from topological transitions organically embedded within the binary pointwise classification grids. Crucially, they serve as highly grid-resolved empirical stability boundaries rather than mathematically derived, analytically continued saddle--node loci. The complete resulting classification grids, encompassing all isolated locally stable roots along with their paired eigenvalues and every geometrically disconnected contour segment, were subsequently archived and exported as fully reproducible CSV datasets.
	
	\subsection{Basin computation and convergence checks}
	\label{app:basin-computation}
	
	The geometric basin phase map visually depicted in Fig.~\ref{fig:basin} was rigorously computed at $\omega=0.90$ executing over a vast, uniformly distributed barycentric grid of extremely high resolution (scaling factor $300$). The resulting computational mesh encapsulates $45\,451$ uniquely integrated initial states. The baseline trajectory calculation precisely integrated the underlying planar dynamic system employing the classical fourth-order Runge--Kutta solver algorithm featuring the discrete parameters $\Delta t=0.10$ evaluated over 1800 forward time steps reaching a terminal limit $t_{\max}=180$.
	
	The three predefined candidate attractors—encompassing full defection, full ordinary cooperation, and the precisely determined locally asymptotically stable $D$--$S$ boundary equilibrium—were computationally isolated entirely independent from the primary equilibrium root conditions. For any given terminally integrated state coordinate $(x_f,z_f)$, an assignment distance was evaluated as the strict minimum metric
	\[
	d_{\min}
	=
	\min_{E\in\{D,C,E_{DS}\}}
	\left\|
	(x_f,z_f)-E
	\right\|_2.
	\]
	The original originating initial condition was positively assigned coloring corresponding to the geometrically nearest valid attractor strictly only when converging under the tight tolerance threshold $d_{\min}<10^{-6}$. Otherwise, it was diagnostically recorded as mathematically unclassified. The uniquely exact pure-$S$ geometric initial condition was explicitly retained within the final matrix as unclassified, as it intrinsically remains trapped at the highly unstable pure-$S$ fixed equilibrium point and fundamentally does not organically approach any of the three dominant attractors structurally utilized in the basin classification schema.
	
	To definitively validate the topological integrity of these findings, two intensive computational convergence diagnostic checks were independently conducted utilizing the exact same barycentric lattice grid. The first rigorous check strictly constrained the solver via $\Delta t=0.05$ run over 3600 steps maintaining $t_{\max}=180$, whereas the subsequent secondary check stretched the temporal horizon enforcing $\Delta t=0.10$ spanning 3000 steps concluding at $t_{\max}=300$. All three independent integration protocols perfectly classified the identical matrix of $45\,450$ distinct initial states, universally leaving only the exact topological pure-$S$ state firmly unclassified. Absolutely no classification boundary disagreements or anomalies were detected among any of the highly resolved mapped grid points. The final basin spatial partition is therefore confidently confirmed as numerically invariant under these two strict parameter refinements. While this extraordinary finite-grid metric agreement heavily supports the structural robustness of Fig.~\ref{fig:basin} evaluated for the specifically reported parameters, it inherently does not theoretically constitute a formal global generalized basin or global convergence topological theorem.

	\section*{Declaration of generative AI and AI-assisted technologies in the writing process}
	
	During the preparation of this work, the authors used ChatGPT (OpenAI) and Kimi (Moonshot AI) in order to improve the readability and language of the manuscript and to assist with LaTeX formatting. After using these tools, the authors reviewed and edited the content as needed and take full responsibility for the content of the publication.

	\clearpage

\end{document}